\documentclass[12pt]{iopart}

\usepackage{graphicx}
\usepackage{epsfig}
\usepackage{verbatim}

\usepackage[T1]{fontenc}
\usepackage{eufrak}

\begin{document}

\newcommand{\be}{\begin{equation}}
\newcommand{\ee}{\end{equation}}
\newcommand{\ba}{\begin{eqnarray}}
\newcommand{\ea}{\end{eqnarray}}
\newcommand{\si}{\sigma_i^z}
\newcommand{\sj}{\sigma_j^z}

\title[Finite temperature behavior of disordered quantum magnets with dissipation] 
{Finite temperature behavior of strongly disordered quantum magnets coupled to a 
dissipative bath}

\author{G. Schehr $^1$ and H. Rieger $^2$}

\address{$^1$ Laboratoire de Physique Th\'eorique, Universit\'e de
  Paris-Sud,  91405 Orsay France \\  
$^2$ Theoretische Physik, Universit\"at des Saarlandes, 66041
  Saarbr\"ucken Germany}   

\begin{abstract}
We study the effect of dissipation on the infinite randomness fixed
point and the Griffiths-McCoy singularities of random transverse Ising
systems in chains, ladders and in two-dimensions. A strong disorder
renormalization group scheme is presented that allows the computation
of the finite temperature behavior of the magnetic susceptibility and
the spin specific heat. In the case of Ohmic dissipation the
susceptibility displays a crossover from Griffiths-McCoy behavior
(with a continuously varying dynamical exponent) to classical Curie
behavior at some temperature $T^*$. The specific heat displays
Griffiths-McCoy singularities over the whole temperature range. For
super-Ohmic dissipation we find an infinite randomness fixed point
within the same universality class as the transverse Ising system
without dissipation. In this case the phase diagram and the parameter
dependence of the dynamical exponent in the Griffiths-McCoy phase can
be determined analytically.
\end{abstract}

\maketitle

\section{Introduction}

The interplay between quantum fluctuations and quenched disorder in
the form of an extensive amount impurities or other random spatial
inhomogeneities can lead to a new class of quantum phase transitions,
governed by an infinite randomness fixed point (IRFP) as established
for transverse Ising models \cite{fisher-irfp} and many other
disordered quantum systems (for an overview see \cite{monthus-igloi}).
Besides unusual scaling laws {\it at} the transition the IRFP is
characterized by a whole parameter range {\it around} the transition,
in which physical observables display singular and even divergent
behavior in spite of a finite spatial correlation length. This is the
manifestation of Griffiths-McCoy singularities or quantum Griffiths
behavior \cite{fisher,qsg-griff,rtic,juhasz,pich}. They have their
origin in rare regions of strongly coupled spins (or other quantum
mechanical degrees of freedoms) that tend to order locally and thus
produce a strong response to small external fields, long relaxation
(or tunneling) times and small excitation energies. 

If the underlying quantum phase transition is governed by an infinite
randomness fixed point the statistics of these rare events leads to a
power law divergence of the susceptibility in a region around the
quantum critical point with a continuously varying exponent. This
dynamical exponent determines all singularities in the Griffiths-McCoy
phase. Continuously varying exponents, interrelated in a specific way
for different physical observables, were observed in many
heavy-fermion materials, and it was argued that this is a
manifestation of Griffiths-McCoy behavior due to an underlying IRFP
\cite{jones,stewart}. In essence these systems form local moments that 
interact via long-range RKKY interaction and have a strong Ising
anisotropy, such that an effective model describing these degrees of
freedom and their interaction is a random transverse Ising system.

Later it was argued that due to the interaction via band electrons the
effective spin degrees of freedom are strongly coupled to a
dissipative Ohmic bath \cite{castro,millis}. From this point of view
the rare regions should be described by spin-boson systems, which are
known to behave classically for sufficiently strong coupling to the
dissipative bath \cite{spin_boson_review}, which would destroy the
expected Griffiths-McCoy singularities.

Since in the presence of dissipation rare regions can undergo phase
transitions and freeze independently from one another (like in the
McCoy-Wu model in the mean-field approximation \cite{palagyi}), the
global phase transition of the system is destroyed by smearing
because different spatial parts of the system 
order at different values of the control parameter \cite{vojta,vojta-rev}.

Recently we analyzed the random transverse Ising chain coupled to a
Ohmic dissipative bath with a strong disorder renormalization group
(SDRG) scheme and could demonstrate that the transition is indeed
smeared, but argued that Griffiths-McCoy singularities are still
observable, at least down to very low temperatures also in the
presence of dissipation \cite{us_prl}. This was done by analyzing the
gap and cluster distribution. In this paper we continue and extend
this SDRG study by a) analyzing the low temperature behavior of the
magnetic susceptibility and the spin specific heat in the case of
Ohmic dissipation, where we will argue that Griffiths-McCoy
singularities are visible at all temperatures in the specific heat and
above a (small) crossover temperature in the susceptibility; b)
considering in addition to chains also ladders and two-dimensional
systems, where we obtain similar results as for the chain; and c)
applying the SDRG also to super-Ohmic dissipation, where we find a
quantum phase transition belonging to the same IRFP universality class
as the system without dissipation and compute analytically the phase diagram
and dynamical exponent in the Griffiths-McCoy phase.

The system that we study is the random transverse Ising model where
each spin is coupled to a dissipative bath of harmonic oscillators,
{\it i.e.}\ ferromagnetically coupled spin-boson systems
\cite{spin_boson_review}. It is defined on $d$-dimensional square 
lattice of linear size $L$ with periodic boundary conditions (pbc) and
described by the Hamiltonian
\begin{eqnarray}
H = - \sum_{\langle i,j\rangle} J_{ij} \sigma_i^z \sigma_j^z - \sum_i
\biggl[
h_i \sigma_{i}^x  + \sum_{k} 
\Bigl(C_{k,i} \hat x_{k,i} \sigma_i^z+
\frac{\hat p_{k,i}^2}{2} + 
\omega_{k,i}^2 \frac{\hat x_{k,i}^2}{2}  
\Bigr)\biggr] \;, 
\label{Def_H} 
\end{eqnarray}
where $\sigma_i^{x,z}$ are Pauli matrices and the masses of the
oscillators are set to one.  
The quenched random bonds $J_i$
(respectively random transverse field $h_i$) are uniformly distributed
between $0$ and $J_0$ (respectively between $0$ and $h_0$). The
properties of the bath are specified by its spectral function ${\cal
  J}_i(\omega)$ 
\begin{eqnarray}
{\cal J}_i(\omega) = \frac{\pi}{2} \sum_{k} \frac
{C_{k,i}^2}{\omega_{k,i}}\delta(\omega - \omega_{k,i}) =
\frac{\pi}{2} \alpha_i \Omega_i^{1-s} \omega^s \theta(\Omega_i -
\omega) \;,
\label{spectral} 
\end{eqnarray}
where $\Omega_i$ is a cutoff frequency and $\theta(x)$ the Heaviside
function such that $\theta(x) = 1$ if $x > 0$ and $\theta(x) = 0$ if $x<0$.
The case $s=1$ is known as Ohmic dissipation although $s>1$
(respectively $s<1$) corresponds to a super-ohmic (respectively
sub-ohmic) dissipation. Initially the spin-bath 
couplings and cut-off frequencies are site-independent, {\it i.e.}
$\alpha_i=\alpha$ and $\Omega_i=\Omega$, but both become
site-dependent under renormalization. 

\section{Real space renormalization.}

\subsection{Decimation procedure.}

In this section, we derive in detail the real space renormalization
scheme to study dissipative random transverse Ising model as in
Eq. (\ref{Def_H}). For simplicity, we present the calculation in
dimension $d=1$ (extensions to higher dimensions are discussed below)
and focus on the random transverse Ising chain (RTFIC):  
\begin{eqnarray}
H_{1d}=\sum^L_{i=1}
\biggl[-J_i \sigma_i^z \sigma_{i+1}^z - h_i \sigma_i^{x}
+ \sum_k \Bigl(C_{k,i} \hat x_{k,i} \sigma_i^z+
\frac{\hat p_{k,i}^2}{2} + 
\omega_{k,i}^2 \frac{\hat x_{k,i}^2}{2}
\Bigr)\biggr] \;.
\label{Def_RTFIC} 
\end{eqnarray}
To characterize the ground state properties of this
system~(\ref{Def_H}), we follow the idea of a real space
renormalization 
group (RG) procedure introduced in Ref. \cite{ma_dasgupta_rsrg} and pushed
further in the context of the RTFIC without dissipation in
Ref. \cite{fisher}. The strategy is to find the largest coupling
in the chain, either a transverse field or a bond, compute the ground
state of the associated part of the Hamiltonian and treat the
remaining couplings in perturbation theory. The bath degrees of
freedom are dealt with in the spirit of the ``adiabatic
renormalization'' introduced in the context of the (single) spin-boson (SB)
model \cite{spin_boson_review}, where it describes
accurately its critical behavior~\cite{vojta_rg_num}.

\subsubsection{When the largest coupling is a bond.}

Suppose that the largest coupling in the chain is a bond, say
$J_2$. The associated part $H_2$ of the full Hamiltonian $H_{1d}$ in
Eq. (\ref{Def_RTFIC})~is 
\begin{eqnarray}
H_2 = & H_2^{(0)} & +  V  \nonumber \\
      & H_2^{(0)} & =  -J_2\sigma_2^z\sigma_3^z+\sum_{i=2,3}\sum_k
 \Bigl(C_{k,i} \hat x_{k,i}\sigma_i^z+
\frac{\hat p_{k,i}^2}{2}+\omega_{k,i}^2 
\frac{\hat x_{k,i}^2}{2} \Bigr) \label{def_h20} 
\label{def_H2}\\
& V & = -h_2 \sigma_2^x -h_3 \sigma_3^x \label{def_v} \; .
\nonumber
\end{eqnarray}
Let us first focus on $H_2^{(0)}$ in Eq. (\ref{def_h20}) and
first introduce the notation for the spin part $|{\mathbf
  S}\rangle \equiv |S_2, S_3 \rangle$, with $S_i = \pm1$ such that 
$\sigma_i^z |{\mathbf S}\rangle = S_i |{\mathbf S}\rangle$ 
for $i=2,3$. Considering now the two baths on site $i = 2,3$ respectively, 
they are composed of a set of 
harmonic oscillators which are labeled by an integer $k$ (which
formally runs from $0$ to $\infty$) and by $i=2,3$. We denote by $|n_{k,i}
\rangle$ the eigenvalues of these harmonic oscillators such that
\begin{eqnarray}
\left(\frac{\hat p_{k,i}^2}{2}+\omega_{k,i}^2 \frac{\hat x_{k,i}^2}{2} \right)
|n_{k,i} \rangle = (n_{k,i}+\frac{1}{2})\omega_{k,i}|n_{k,i} \rangle
\; .
\label{state}
\end{eqnarray}
In the absence of the coupling between the spins and the bath, 
$H_2^{(0)}$ can be straightforwardly diagonalized by tensorial products
of $|{\mathbf S}\rangle$ and $|n_{k,i} \rangle$. The corresponding eigenvalues
are simply the sum of the eigenvalues of the individual Hamiltonians
in Eq. (\ref{def_h20}) without the last term of interaction. The
coupling between the spins and the baths does not change these
eigenvalues (up to a global shift)
and only affects the eigenstates. We introduce the shifted
"eigenvectors" 
\begin{eqnarray}
|n_{k,j}^{\pm} \rangle = \exp{\left(\pm i
 \frac{C_{k,j}}{\omega_{k,j}^2}\hat p_{k,j}\right)} |n_{k,j} \rangle \quad,
 \quad j=2,3 \; ,
\label{shifted}
\end{eqnarray}
from which we can construct the eigenvectors of $H_2^{(0)}$,
including the interaction between the baths and the spins as
\begin{eqnarray}
|{\mathbf S},{\mathbf n} \rangle &=& |S_2,S_3\rangle \otimes |{\mathfrak
 n}_2^{S_2}\rangle \otimes |{\mathfrak n}_3^{S_3} \rangle \nonumber \\
|{\mathfrak n}_i^{S_i}\rangle &=& \bigotimes_k  |n_{k,i}^{S_i} \rangle \;.
\end{eqnarray}
The eigenvalues of $H_2^{(0)}$ are given by
\begin{eqnarray}
H_2^{(0)} |{\mathbf S},{\mathbf n} \rangle 
&=& E^{(0)}_{{\mathbf S},{\mathbf n}} |{\mathbf S},{\mathbf n} \rangle\;,\nonumber\\
E^{(0)}_{{\mathbf S},{\mathbf n}} 
&=& -J_2 S_2 S_3 + \sum_{i=2,3} \sum_k 
  \biggl[\Bigl(n_{k,i} + \frac{1}{2} \Bigr) \omega_{k,i}
  -\frac{1}{2} \frac{C_{k,i}^2}{\omega_{k,i}^2}\biggr]\;.
\end{eqnarray}
Each level is thus a priori degenerated twice (except accidental
degeneracy) and in the limit of large coupling $J_2$ we first restrict
ourselves to the lowest energy levels, such that $S_2 S_3 =
+1$. Performing perturbation theory in $V$, one obtains that the first
order corrections vanish. To second order in $V$, one has to
diagonalize the $2 \times 2$ matrix $V^{(2)}$ in the eigensubspace
associated to the zeroth order eigenvalue 
$E^{(0)}_{{\mathbf S},{\mathbf n}}$ with $S_2 S_3 = +1$ which is formally given~by
\begin{eqnarray}
V^{(2)} =  \sum_{ {\mathbf S'},{\mathbf n'} ,   E_{{\mathbf
S'},{\mathbf n'}}\neq E_{{\mathbf S},{\mathbf n}}}  
\frac{V|{\mathbf S}',{\mathbf n}'\rangle \langle{\mathbf
    S}',{\mathbf n}'|V}{E^{(0)}_{{\mathbf S},{\mathbf n}} -
  E^{(0)}_{{\mathbf S'},{\mathbf n}'}  }\; . \label{gen_formula_perturbation}
\end{eqnarray}
One obtains from (\ref{gen_formula_perturbation}) the diagonal
elements 
\begin{eqnarray}
V^{(2)}_{11} =V^{(2)}_{22} & = h_2^2 \sum_{{\mathfrak n}'_2} \frac{| \langle
  {\mathfrak n}_2^+ | {{\mathfrak n}'_2}^-\rangle|^2}{-2J_2 +
  \sum_{k} (n_{k,2} - 
  n'_{k,2})\omega_{k,2} } \nonumber \\
& + h_3^2 \sum_{{\mathfrak n}'_3} \frac{| \langle
  {\mathfrak n}_3^+ | {{\mathfrak n}'_3}^-\rangle|^2}{-2J_2 +
  \sum_{k} (n_{k,3} - 
  n'_{k,3})\omega_{k,3} } \;,
\label{diag}
\end{eqnarray}
and the off-diagonal elements 
\begin{eqnarray}
V_{12}^{(2)} = V_{21}^{(2)} = -\frac{h_2 h_3}{J_2} \langle {\mathfrak
  n}_2^-|{\mathfrak n}_2^+  \rangle \langle {\mathfrak
  n}_3^-|{\mathfrak n}_3^+  \rangle \;.
\label{nondiag}
\end{eqnarray}
In the absence of a coupling to the dissipative bath ({\it i.e.}\
$C_{k,i}=0$ for all $k$ and $i$) the shifted eigenstates
(\ref{shifted}) are identical with the non-shifted eigenstates and
therefore $\langle{\mathfrak n}_i^-|{\mathfrak n}_j^+  \rangle=
\delta_{{\mathfrak n}_i,{\mathfrak n}_j}$ and thus $V^{(2)}_{11}
=V^{(2)}_{22} = 
\frac{h_1^2+h_2^2}{-2J_2}$ and $V_{12}^{(2)} = V_{21}^{(2)} = -h_2 h_3/J_2$.
This matrix has two eigenvalues whose difference, the gap, is
$2h_2h_3/J_2$.  Thus for each oscillator state the low lying
excitations of $H_2$ in (\ref{def_H2}), with $S_2=S_3$, can again be
described by an effective two-state system, {\it i.e.}\ a spin in a
transverse field of strength $h'=h_2h_3/J_2$. The spirit of the strong
disorder renormalization group is to keep this effective two-level system
(for each oscillator state) and to neglect the large energy doublet
with $S_2\ne S_3$. In this way one has replaced two spins (with
moments $\mu_2$ and $\mu_3$) and a large coupling $J_2$ between them
by a single effective spin with moment $\mu_2+\mu_3$ in a
small transverse field $h_2 h_3/J_2$, thus one degree of freedom 
with a large energy has been decimated.

In the presence of non-vanishing couplings $C_{k,i}$ to the
oscillators one needs to decimate also the high energy modes of the
bath such that $\omega_{k,i} > pJ_2$, where $p$ is some (large)
number. Given that $J_2$ is a large energy scale, the low lying energy
levels will be those with $n_{k,i} = 0$ for $\omega_{k,i} >
pJ_2$. Therefore we decompose the oscillator states according to
\begin{equation}
| {\mathfrak n}_i^S\rangle
=|{\mathfrak n}_i^{<S}\rangle \otimes
|{\mathfrak n}_i^{>S}\rangle\;,
\end{equation}
with $S=\pm1$ and
\begin{equation}
|{\mathfrak n}_i^{<S}\rangle=
\bigotimes_{k_i^<}
|n_{k_i}^{S} \rangle
\quad{\rm and}\quad
|{\mathfrak n}_i^{>S}\rangle=
\bigotimes_{k_i^>}
|n_{k_i}^{S} \rangle\;,
\end{equation}
where $k_i^<={\{k|\omega_{k,i}\le pJ_2\}}$ and 
$k_i^>={\{k|\omega_{k,i}> pJ_2\}}$.
Additionally we introduce the product state of oscillators which 
are in the ground state by
$|{{\mathfrak 0}_i^{>}}^{S_i}\rangle =
\bigotimes_{k_i^>} |0_{k_i}^{S_i} \rangle$. At energy scales smaller
than $pJ_2$ all oscillators with frequencies larger than $pJ_2$ will
be in their ground states, and therefore we will consider the matrix
elements in (\ref{diag}) and (\ref{nondiag}) only for oscillator states
$|{\mathfrak n}^+\rangle= |{\mathfrak
n}^{<+}\rangle\otimes|{\mathfrak o}^{>+}\rangle$.
For these states the two sums on the r.h.s.\ of (\ref{diag}) read
\begin{equation}
\sum_{{\mathfrak m}_i^<,{\mathfrak m}_i^>} 
h_i^2\frac{|\langle{\mathfrak n}_i^{<+}|{\mathfrak m}_i^{<-}\rangle|^2\cdot
      |\langle{\mathfrak o}_i^{>+}|{\mathfrak m}_i^{>-}\rangle|^2}
{-2J_2-\sum_{k^>} m_{k,2}\omega_{k,2}+\sum_{k^<} (n_{k,2}-m_{k,2})\omega_{k,2}}\;.
\end{equation}
To leading order in $J_2$ one can neglect the term $\sum_{k^<}
(n_{k,2}-m_{k,2})\omega_{k,2}$, since it involves only frequencies
smaller than $p J_2$. Then the sum over the low frequency oscillator
states $|{\mathfrak m}_i^{<}\rangle$ yields one since they form a
complete basis for the low frequency oscillator Hilbert space and the
individual terms in the sum do not depend on the quantum numbers
${\mathfrak n}_i^<$ any more.  Thus the diagonal matrix elements in
(\ref{diag}) read to leading order in $J_2$
\begin{eqnarray}
&&V^{(2)}_{ii}
= h_2^2 \sum_{{\mathfrak m}_2^{>}}
\frac{|\langle {\mathfrak 0}_2^{>+} 
|{{\mathfrak m}_2^{>}}^- \rangle |^2}{-2J_2  -
  \sum_{k_2^>} m_{k_2}\omega_{k,2}} 
+ h_3^2  \sum_{{\mathfrak m}_3^{>}}\frac{|\langle {\mathfrak 0}_3^{>+} 
  |{{\mathfrak m}_3^{>}}^- \rangle |^2}{-2J_2  - \sum_{k_3^>} m_{k_3}\omega_{k,3}}\;.
\end{eqnarray}
Note that this expression does not depend on the quantum numbers
${\mathbf n}^{<}$ for the low frequency oscillators.
For the non-diagonal matrix elements in (\ref{nondiag}) one gets
\begin{eqnarray}
&&V_{12}^{(2)} = V_{21}^{(2)} = - \frac{{\mathcal A}h_2 h_3}{J_2}
  \langle {\mathfrak n}_2^{<-}|{\mathfrak n}_2^{<+}  \rangle 
  \langle {\mathfrak n}_3^{<-}|{\mathfrak n}_3^{<+}  \rangle 
\end{eqnarray}
with
\begin{eqnarray}
&&{\mathcal A} =
  \langle {\mathfrak 0}_2^{>+} | {\mathfrak 0}_2^{>-}\rangle 
  \langle {\mathfrak 0}_3^{>+} | {\mathfrak 0}_3^{>-}\rangle 
= \prod_{k_2^>}\langle 0_{k_2}^{+}  |0_{k_2}^{-} \rangle  
  \prod_{k_3^>}\langle 0_{k_3}^{+}  |0_{k_3}^{-} \rangle  \;.
\end{eqnarray}
The amplitude ${\mathcal A}$ can be then expressed in terms of the
spectral density, using that $\langle 0_{k,i}^{+}  |0_{k,i}^{-}
\rangle = \exp{-(C_{k,i}^2/\omega_{k,i}^3)}$. This yields
\ba
{\mathcal A} = \exp{\left(-\frac{2}{\pi}\int_{pJ_2}^{ \Omega_2}
  \frac{J_2(\nu)}{\nu^2} d\nu - \frac{2}{\pi}\int_{pJ_2}^{
    \Omega_3} \frac{J_3(\nu)}{\nu^2} d\nu    \right)} \;,
\label{eq_amplitudeA} 
\ea
where we have used the definition of the spectral density in
Eq. (\ref{spectral}). Since the 
diagonal term does not depend on ${\mathbf n}^{<}$ 
the diagonalization of $V^{(2)}$ yields (up to second order)
the following correction to the lowest eigenvalues 
\begin{eqnarray}
E^{(2)}_{\pm,{\mathbf n}^<} = {\cal E}_0 + \sum_{i=2,3} \sum_k
\left(n_{k,i}+\frac{1}{2}\right)\omega_{k,i} \pm
\frac{{\mathcal A}h_2 h_3}{J_2} 
  \langle {\mathfrak 
  n}_2^{<-}|{\mathfrak n}_2^{<+}  \rangle \langle {\mathfrak
  n}_3^{<-}|{\mathfrak n}_3^{<+}\rangle \;,\nonumber \label{low_level_H2}\\
\end{eqnarray}
where ${\cal E}_0 = -J_2 + V_{11} -\frac{1}{2} \sum_{i=2,3} \sum_k
C_{k,i}^2/\omega_{k,i}^2$ is a constant, independent of 
${\mathbf n}^{<}$. We now consider an effective spin-boson
Hamiltonian coupled to {\it both} baths $2$ and $3$:
\ba
\tilde H_2 = -\tilde h_2 \sigma_{2}^x +  \sum_{i=2,3} \sum_{k} \biggl( \frac{\hat
  p_{k,i}^2}{2} +  
\omega_{k,i} \frac{\hat x_{k,i}^2}{2} + C_{k,i} \hat
x_{k,i} \sigma_2^z  \biggr)
\ea 
where the frequencies are such that $\omega_{k,i} < pJ_2$. The
effective spins being coupled to both baths, one has
\ba
\tilde {\cal J}_2(\omega) = \theta(pJ_2 - \omega) ({\cal J}_2(\omega)
+ {\cal J}_3(\omega)) \;.
\label{sum_of_bath} 
\ea
Treating the small
parameter $\tilde h_2$ in (degenerate) perturbation theory, one obtains
the low lying eigenvalues of $\tilde H_2$ to first order in $\tilde h_2$:
\begin{eqnarray}
\tilde E^{(1)}_{\pm,{\mathbf n}^<} = \sum_{i=2,3} \sum_k
\left(n_{k,i}+\frac{1}{2}\right)\omega_{k,i} \pm \tilde h_2 
  \langle {\mathfrak 
  n}_2^{<-}|{\mathfrak n}_2^{<+}  \rangle \langle {\mathfrak
  n}_3^{<-}|{\mathfrak n}_3^{<+}\rangle \;.\label{low_level_SB}
\end{eqnarray}
The comparison between Eq. (\ref{low_level_H2}) and
Eq. (\ref{low_level_SB}) shows that the low energy spectrum of the two
interacting spin-bosons in $H_2$ can be described by a single
spin-boson system with renormalized parameters given by
\begin{eqnarray}
&&\tilde h_2 = \frac{{\mathcal A}h_2 h_3}{J_2}  \label{decim1_j_ohm} \\
&&\tilde \alpha_2 = \alpha_2 + \alpha_3 \quad, \quad \tilde \Omega_2 =
  pJ_2 \;, \label{decim2_j_ohm}
\end{eqnarray}
where ${\mathcal A}$, which depends on the parameters of the
Hamiltonian $H_2$, is given by Eq. (\ref{eq_amplitudeA}) and the
equalities in (\ref{decim2_j_ohm}) are a direct consequence of
Eq. (\ref{sum_of_bath}). This effective spin-boson will interact
ferromagnetically with the spin-boson on site $1$ and site $4$ with
couplings
\ba
\tilde J_1 = J_1 \quad, \quad \tilde J_2 = J_3  \;. \label{decim3_j_ohm}
\ea
These relations (\ref {decim1_j_ohm}),  (\ref {decim2_j_ohm}) and
(\ref {decim3_j_ohm}) constitute the first  
set of decimation rules.

\subsubsection{When the largest coupling is a field.}

Suppose now that the largest coupling in the chain is a transverse field,
say $h_2$. Before we treat the coupling of site $2$ to the rest of
the system $-J_1\sigma_1^z\sigma_2^z-J_2\sigma_2^z\sigma_3^z$ 
perturbatively as in \cite{fisher} we consider
the part of the Hamiltonian
that represents a single spin-boson system
\ba
H_2' = -h_2\sigma_2^x+\sum_k (\frac{\hat p_{k,2}^2}{2}+\omega_{k,2} \frac{\hat
  x_{k,2}^2}{2} + C_{k,2} \hat x_{k,2}\sigma_2^z) \;.
\ea
In this case, one would like
to have a way to decimate the high energy modes of the bath, here
the harmonic oscillators such that $\omega_{k,2} > ph_2$, where $p$ is
some (large) number. 
Since for those oscillators $\omega_{k,2} \gg h_2$ one
can assume 
that they adjust instantaneously to the current value of $\sigma_2^z$
the renormalized energy splitting is easily calculated using
Eq. (\ref{low_level_SB}) -- the so called adiabatic renormalization
\cite{spin_boson_review}-- 
and one gets an effective transverse field
$\tilde h_2<h_2$: 
\begin{eqnarray}
&&\tilde h_2 = {\mathcal A}' h_2 \;, \;  \tilde \Omega_2 =
  p h_2  \label{decim1_h_ohm} \\ 
&&{\mathcal A}' =
  \exp{\left(-\frac{2}{\pi}\int_{ph_2}^{ \Omega_2}   
  \frac{{\cal J}_2(\nu)}{\nu^2} d\nu \right)} \;.  \label{eq_amplitudeAprime}
\end{eqnarray}
If $\tilde h_2$ is still the largest
coupling in the chain the iteration (\ref{decim1_h_ohm}) is repeated.
Two situations may occur depending on the parameters $s$ and
$\alpha_i$. 
\begin{itemize}
\item{
If $s > 1$ or $s=1$ and $\alpha_2 < 1$ this procedure
(\ref{decim1_h_ohm}) will converge to a finite value $h_2^*$ given by 
\ba
\hspace*{-1cm}h_2^* = h_2 \exp{\left(-\frac{2}{\pi}\int_{ph_2^*}^{ \Omega_2}  
  \frac{{\cal J}_2(\nu)}{\nu^2} d\nu \right)} \sim
\cases{
h_2 \exp{\left(-\frac{\alpha_2}{s-1}\right)} \quad, \quad s>1\\ 
h_2\left(\frac{ph_2}{\Omega_2}\right)^{\frac{\alpha_2}{1-\alpha_2}} \quad,
\quad s=1, \alpha_2<1 \; ,
} \label{h_2star_explicit}
\ea
where the expression of $h_2^*$ for $s>1$ is valid only in the limit
$ph_2 \ll \Omega_2$. In this case, the spin-boson system at site $2$ is in a
delocalized phase 
in which the spin and the bath can be considered as being decoupled
(formally $\alpha_2=0$), as demonstrated by an RG treatment 
in \cite{vojta_rg_num}. If this value $h_2^*$ (\ref{h_2star_explicit})
  is still the largest 
coupling in the 
chain the spin on site $2$ will be aligned with the transverse
field. As in the RTFIC 
without dissipation, this spin is then decimated (as it will not
contribute to the magnetic susceptibility) and gives rise, in second
order degenerate perturbation theory, to an effective coupling
$\tilde{J_1}$ between the neighboring moments at site $1$ and
$3$~\cite{fisher}
\begin{eqnarray}
\tilde J_1 = \frac{J_1 J_2}{h_2^*} \;. \label{decim2_h_ohm}
\end{eqnarray}
}
\item{
If $s=1$ and $\alpha_2 >1$, $\tilde h_2$ can be made arbitrarily small by
repeating the procedure (\ref{decim1_h_ohm}) implying that the 
SB system on site $2$ is in its localized phase 
\cite{vojta_rg_num} and essentially behaves classically:
the decimation rule (\ref{decim1_h_ohm}) indeed amounts to set
$\tilde{h}_2=0$. Such a moment, or cluster of spins, will be aligned
with an infinitesimal external longitudinal field and is
denoted as ``frozen''.}
\end{itemize}
These relations (\ref{decim1_h_ohm}-\ref{decim2_h_ohm}) constitute the second set 
of decimation rules. The complete decimation procedure is sketched 
for the Ohmic case in Fig. 1.

\begin{figure}[h]
\begin{minipage}{0.5\linewidth}
\includegraphics[angle=270,width=\linewidth]{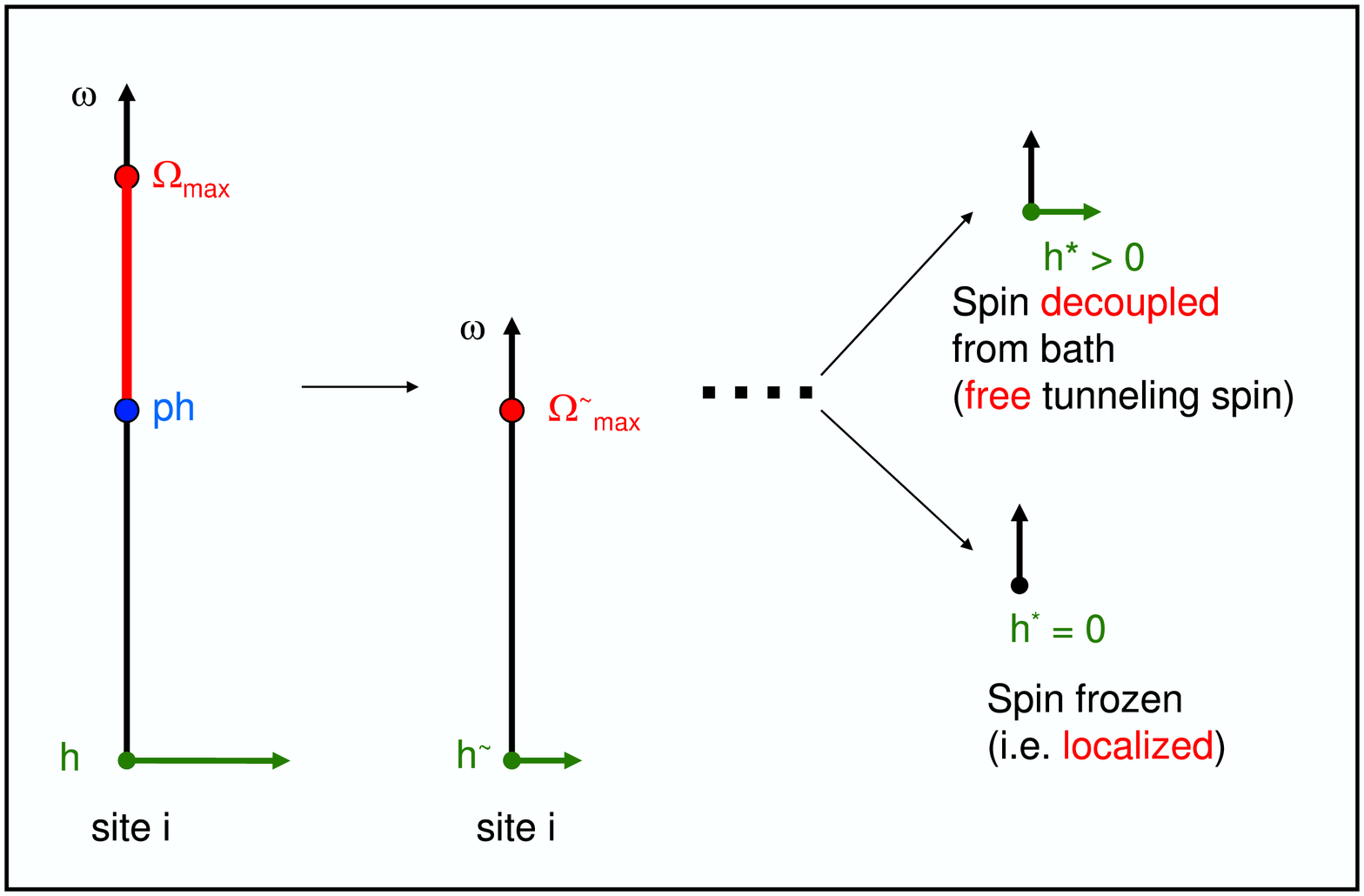}
\end{minipage}\hfill
\begin{minipage}{0.5\linewidth}
\vspace*{-0.4cm}
\includegraphics[angle=270,width=\linewidth]{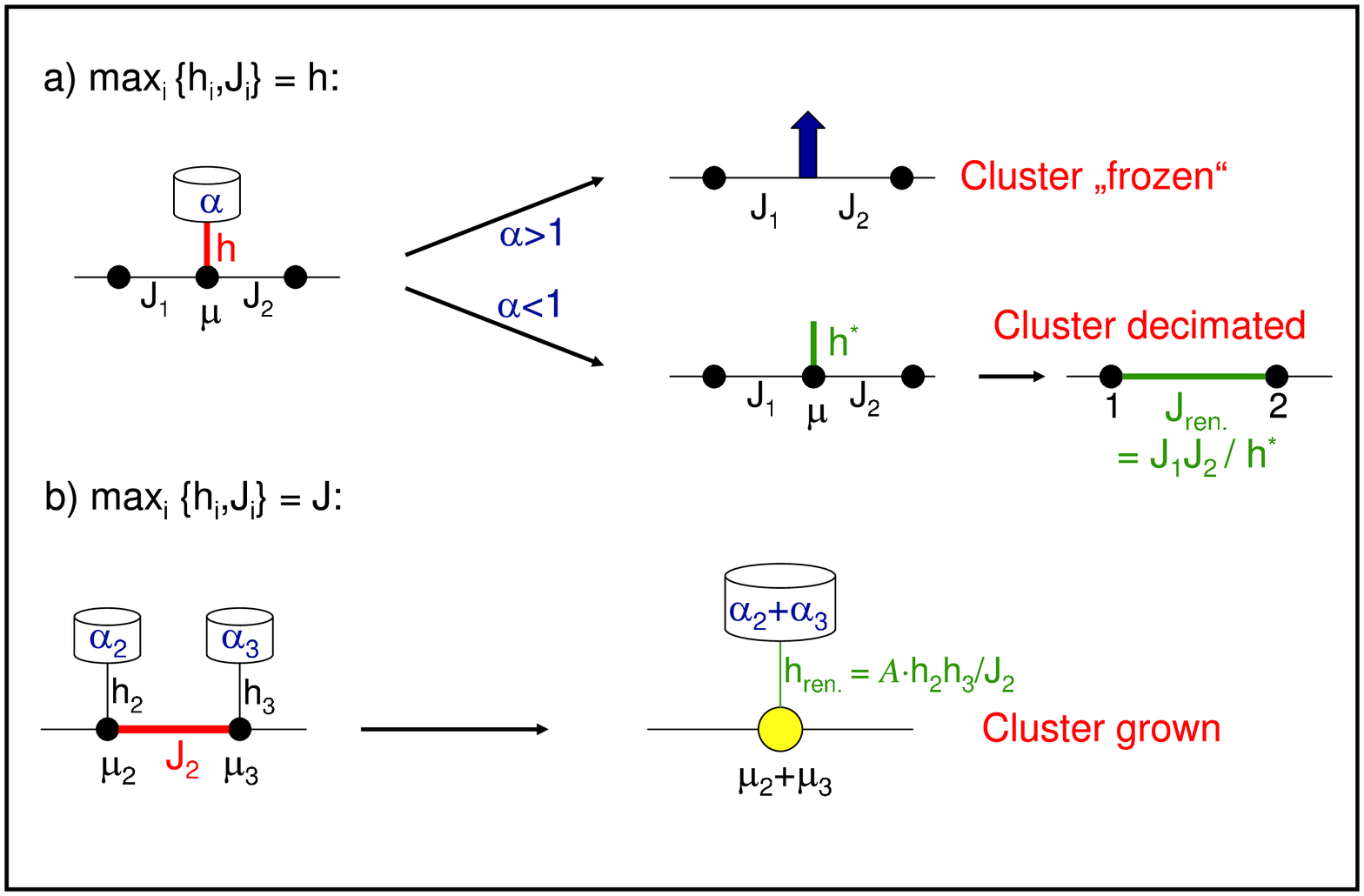}
\end{minipage}
\caption{{\bf Left:} Sketch of the adiabatic renormalization of the
bosonic bath for a single spin-boson system in the Ohmic case: The
fast oscillators with frequencies $\omega>ph$, where $h$ is the
transverse field acting on the spin and $p$ an arbitrary large
parameter, are treated in the adiabatic approximation resulting
in a renormalized transverse field $\tilde{h}$ for the spin 
and a decreased upper cut-off frequency $\tilde{\Omega}_{\rm max}$
for the bosonic bath. This procedure is continued until it runs
into a fixed point, where either the renormalized field $h^*$ vanishes
(for $\alpha>1$) and the spin is frozen, or the renormalized field $h^*$ 
has a positive value, in which case it is a nonfrozen spin in a 
transverse field $h^*$, decoupled from the bath.
{\bf Right:} Sketch of the SDRG in the presence 
of a dissipative bath for the Ohmic case. When a field $h$ is the largest 
coupling (top), first the oscillator bath is renormalized along the
lines sketch 
on the left site. Once this is done and the field is still the largest coupling
the spin gets decimated as described in the text. When a bond $J$ between two
spins at site 2 and 3 is the largest coupling the two spins get decimated to 
form a cluster with moment $\mu_2+\mu_3$, coupled to a combination of bath 
2 and 3, {\it i.e.} a new dissipative bath with coupling strength $\alpha_2+\alpha_3$.}
\label{Sketch}  
\end{figure}

\subsection{Numerical implementation.}

In the following we analyze this RG procedure defined by the
decimation rules (\ref{decim1_j_ohm}-\ref{decim3_j_ohm}, 
\ref{decim1_h_ohm}-\ref{decim2_h_ohm}) numerically. This
is done by considering a finite system of linear size $L$ with 
pbc and iterating the decimation rules until only one site is left. This
numerical implementation has been widely used in previous works
\cite{monthus-igloi, yu-cheng_statphys}, and it has been shown to reproduce 
with good accuracy the exact results of Ref. \cite{fisher} for the
RTFIC \cite{yu-cheng_statphys}. In particular, the transverse field
$h$ acting on the last remaining spin is, at low ferromagnetic
coupling $J_0$, an estimate for the smallest excitation energy. Its
distribution, $P_L(h/\Gamma_0)$, where $\Gamma_0$ is the largest
coupling in the initial system of linear size $L$, reflects the
characteristics of the gap distribution \cite{rtic}. This
quantity, and specifically its dependence on the system size $L$ can
be efficiently used to characterize Griffith-Mc Coy singularities and
critical behavior characterized by an infinite randomness fixed point.

\section{Ohmic dissipation.}

Ohmic dissipation means $s=1$ in (\ref {spectral}), {\it i.e.} a spectral
function for the oscillators that is linear in the frequency (up to
the upper cut-off $\Omega$). For a
single spin in a transverse field $h$ and coupled to such an Ohmic
bath, a lot of results are 
available \cite{spin_boson_review}.  Here we mention only that this
system has a phase transition at zero temperature driven by the
coupling strength $\alpha$. For small $\alpha$ the spin can still
tunnel quantum mechanically, whereas for large $\alpha$ the spin is
frozen and behaves classically, the critical coupling strength
$\alpha_c(h)$ is equal to $1$ in the limit where $h/\Omega \ll 1$
where $\Omega$ is the cut-off frequency of the bath, an exact result 
predicted correctly by the adiabatic
approximation mentioned above. Such a transition
is also present in an infinite ferromagnetic spin chain coupled to a
dissipative bath, as it was shown recently numerically
\cite{troyer}. Here we want to focus on the interplay of disorder, quantum
fluctuations and dissipation and study random transverse field
Ising systems coupled to a dissipative environment by implementing the
decimation rules 
(\ref{decim1_j_ohm}-\ref{decim3_j_ohm},
\ref{decim1_h_ohm}-\ref{decim2_h_ohm}) for ohmic dissipation. For
$s=1$, the amplitudes ${\cal A}$ in Eq. (\ref{eq_amplitudeA}) and ${\cal A}'$
in Eq. (\ref{eq_amplitudeAprime}) which enter these decimation rules are
given by
\ba
{\cal A} = \left(\frac{p J_2}{{\Omega}_2}
\right)^{\alpha_2}  \left(\frac{p J_2}{{\Omega}_3}
\right)^{\alpha_3} \;, \; {\cal A}' = \left( \frac{p h_2}{\Omega_2}
\right)^{\alpha_2} \;.
\ea

\subsection{One dimensional system : Random transverse field Ising chain.}

\subsubsection{Gap distribution : finite size analysis.}

The RTFIC coupled to a ohmic bath was treated in detail in
Ref. \cite{us_prl}. We just recall here the main results. Since the
last spin can either be frozen  
(i.e the last field $h$ is zero) or non-frozen we split 
$P_L(h/\Gamma_0)$ into two parts:
\begin{eqnarray}
P_L(h/\Gamma_0) = A_L \tilde P_L(h/\Gamma_0) + (1-A_L)
\delta(h/\Gamma_0) \;, \label{gen_form} 
\end{eqnarray}
where $\tilde P_L(h/\Gamma_0)$ is the restricted distribution of the
last fields in the samples that are non-frozen and $A_L$ is the
fraction of these samples. It, or equivalently $\tilde
P_L(\log(\Gamma_0/h))$, represents the distribution of the
smallest excitation energy in the ensemble of non-localized spins. At
low coupling (small $J_0$ or small $\alpha$), $\tilde
P_L(\log(\Gamma_0/h))$ shows indications of Griffiths-McCoy
singularities characterized by the 
following scaling behavior for $\tilde P_L$
\begin{eqnarray}
\tilde P_L(\log(\Gamma_0/h)) = {\cal P}(\log(\Gamma_0/h L^z))\;, 
\label{griffith_scaling}
\end{eqnarray}
where $z$ is a dynamical exponent continuously varying with ($J_0$,
$\alpha$, etc.). As the coupling is increased, $z$ is also increasing and
eventually, at some pseudo-critical point, $\tilde
P_L(\log(\Gamma_0/h))$ exhibits a scaling which is characteristic for an 
IRFP:
\begin{eqnarray}
\tilde P_L(\log(\Gamma_0/h)) = 
L^{-\psi}{\cal P}_{\rm{IRFP}}(L^{-\psi}\log(\Gamma_0/h))
\label{irfp_scaling} 
\end{eqnarray}
whith $\psi \simeq 0.32$ \cite{us_prl} is a critical exponent
characterizing the 
IRFP. Notice that this value of $\psi$ is different from $\psi_{\rm
  RTFIC} = 1/2$ computed 
exactly for the RTFIC \cite{fisher}. The main striking point in the
case of ohmic dissipation is that 
although the restricted distribution $\tilde P_L(\log(\Gamma_0/h))$
displays Griffith's like behavior like in Eq. (\ref{griffith_scaling})
the magnetization becomes {\it finite} above a certain length scale
$L^*$. This finite magnetization 
is a manifestation of the ``frozen'' clusters which lead to the
concept of rounded quantum phase transitions in the presence of
dissipation \cite{vojta}. Due to these ``frozen'' clusters, the
amplitude $A_L$ decays exponentially above $L^*$, with $A_L
\propto e^{-L/L^*}$.  
However, as we pointed out in Ref. \cite{us_prl}, the interpretation
of the finite size analysis
(\ref{gen_form},\ref{griffith_scaling},\ref{irfp_scaling}) in presence
of dissipation has to be done carefully. Indeed in Ref. \cite{us_prl}
we suggested that despite the presence of these frozen clusters,
Griffith's singularities should be observable in the susceptibility
$\chi(T)$, above a certain temperature $T^* \propto L^{*-z}$ as well
as in the specific heat $C_v(T)$. This property can actually be shown
(see \ref{rtic_diluted}) on a toy model where one considers a RTFIC
without dissipation but with a finite fraction $\rho$ of zero
transverse fields.

Here we will use this strong disorder approach to extract
thermodynamical properties of the full problem described by the
Hamiltonian (\ref{Def_RTFIC}).

\subsubsection{Susceptibility at finite temperatures.}
The SDRG successively eliminates degrees of freedom with a large
excitation energy from the starting Hamiltonian, thereby reducing
continuously the maximum energy scale of the effective Hamiltonian. If
continued down to the smallest energy scale the final effective
Hamiltonian (consisting only of a single but large cluster in an
effective transverse field) provides information about the ground
state of the starting spin chain, the gap, the size, the geometry,
{\it etc...} of
the smallest excitation energy. To extract information on
thermodynamic properties, at low but non-vanishing temperatures, one
has to stop the RG procedure at an energy scale of the same order of
magnitude as the temperature $T$: clusters (or degrees of freedom)
that are already frozen at this energy scale will not be active at
this temperature and behave like classical spins (at this temperature). 
The thermodynamical properties, observables like susceptibility or specific 
heat, will be determined by the active, {\it i.e.} not yet frozen clusters.

\begin{figure}[h]
\begin{minipage}{0.5\linewidth}
\includegraphics[angle=0,width=\linewidth]{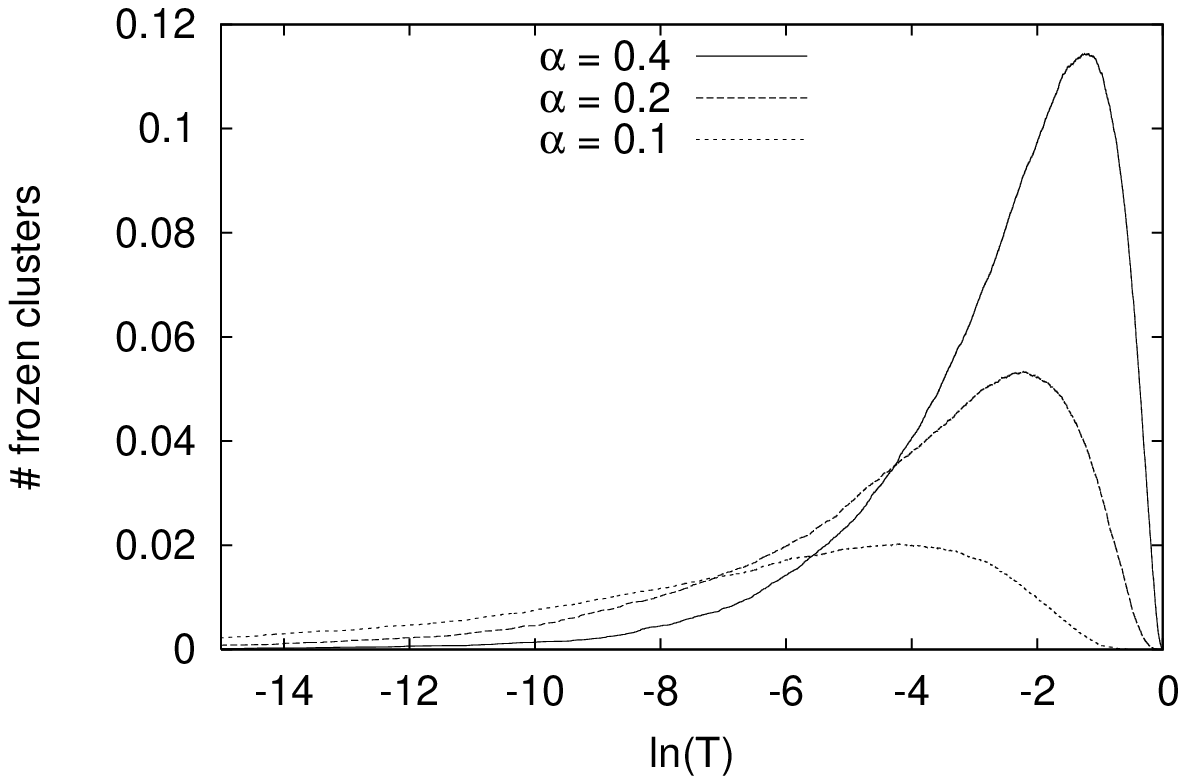}
\end{minipage}\hfill
\begin{minipage}{0.5\linewidth}
\includegraphics[angle=0,width=\linewidth]{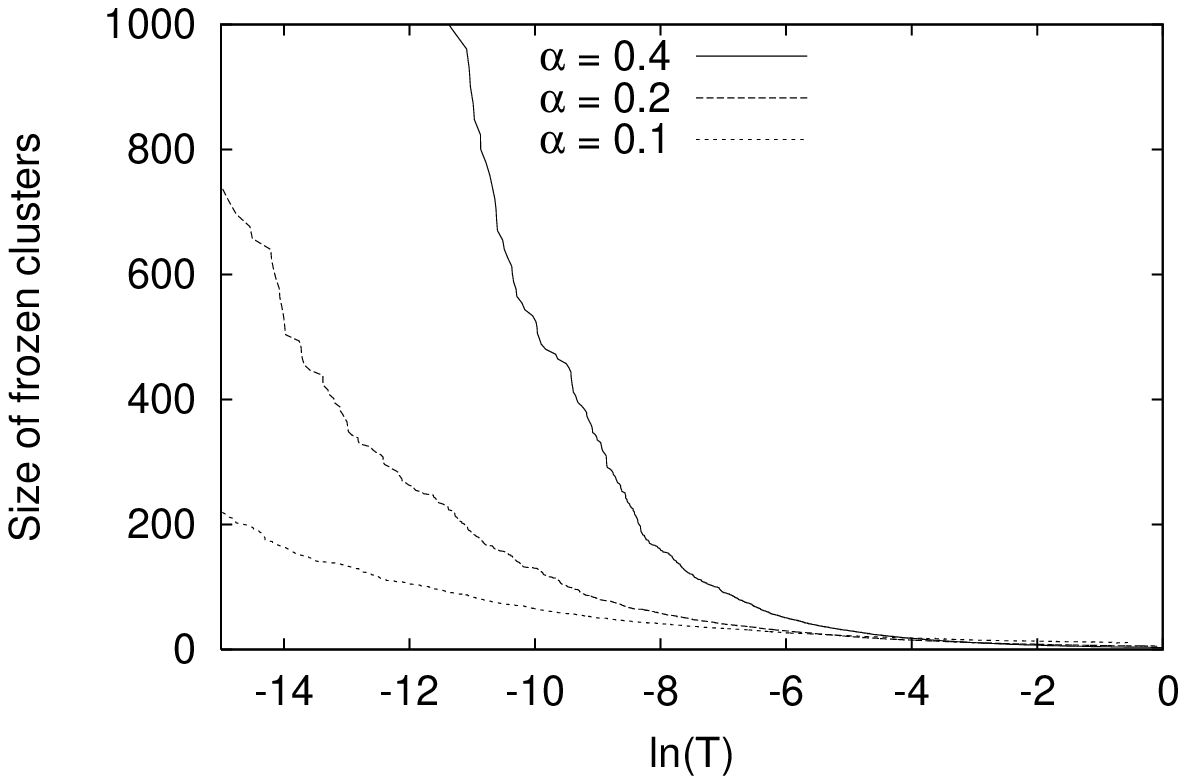}
\end{minipage}
\caption{
  Number (left) and size (right) of frozen clusters
  in the disordered chain coupled to an ohmic bath
  as a function of temperature $T$ for different values of
  $\alpha$. Shown are data for a single large disorder realization,
  the size is $L=40000$, the disorder strength is $h_0 = 1$, $J_0 =
  0.025$. For this value of $J_0$ the pseudo
  critical point is located at $\alpha=0.2$ \cite{us_prl}. 
  \label{number_size_frozens}}
\end{figure}

It is instructive to have a look at the number and size of frozen
clusters as a function of the upper cut-off energy, which we identify
now with the temperature $T$. As one can see from the left panel of
Fig. \ref{number_size_frozens} the number of frozen clusters is zero
at high temperatures (simply because $\alpha<1$) and increases rapidly
with decreasing temperature before it reaches a maximum and then
decreases. The initial increase is due to the formation of many small
clusters that behave like classical spins at the corresponding
temperature with moments of the order of $10$. The subsequent decrease
of the number of clusters correlates with an increase in the size of
the clusters 
as can be seen in th right panel of Fig. \ref{number_size_frozens}
and which is due to the coalescence of small clusters into larger ones
at the corresponding temperatures. 

\begin{figure}[h]
\begin{minipage}{0.5 \linewidth}
\begin{center}
\includegraphics[angle=-90,width=\linewidth]{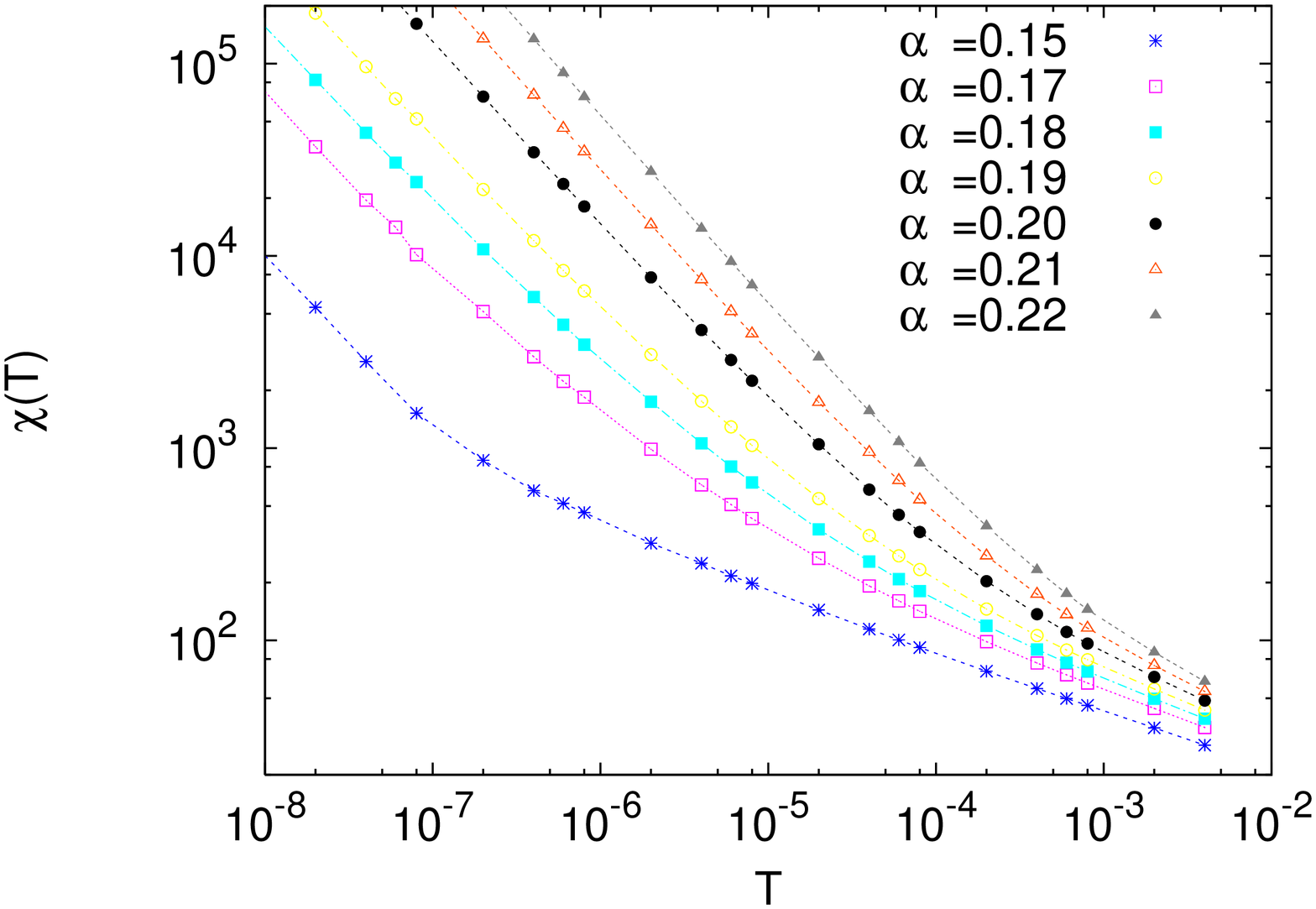}
\end{center}
\end{minipage}\hfill
\begin{minipage}{0.5 \linewidth}
\begin{center}
\includegraphics[angle=-90,width=\linewidth]{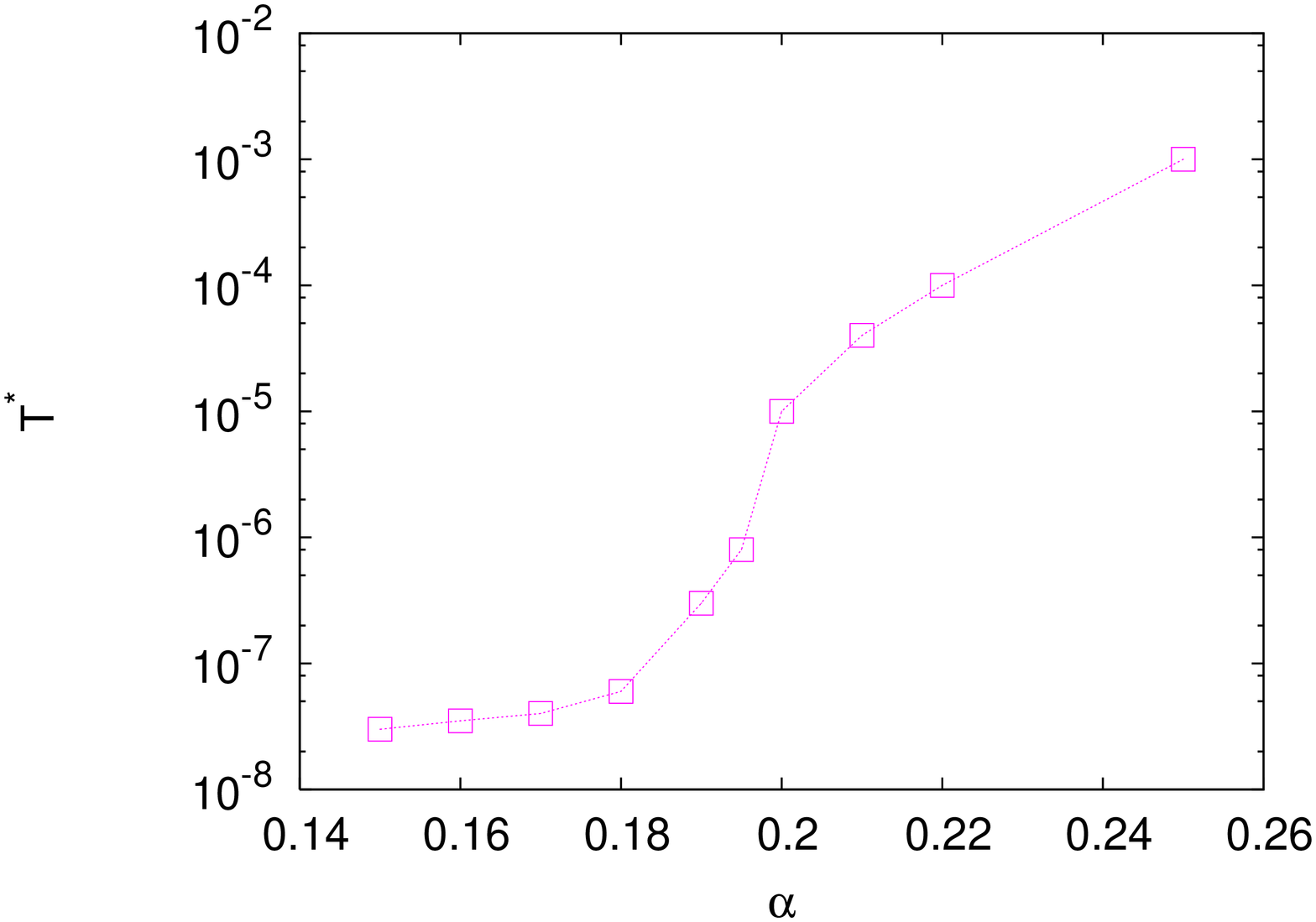}
\end{center}
\end{minipage}
\caption{{\bf Left}: Magnetic susceptibility $\chi(T)$ for a disordered chain
  coupled to an ohmic bath as a function of temperature for different
  coupling strength $\alpha$. The susceptibility is calculated as
  $\chi(T)= \chi_{\rm active}(T) + \chi_{\rm frozen}(T)$ using
  Eq. (\ref{suscept_nonfrozen}) and Eq. (\ref{suscept_frozen}).  The
  size is $L=4096$, the disorder strength is $h_0=1$, $J_0 =
  0.025$. For this value of $J_0$ the pseudo-critical point is located
  at $\alpha=0.2$ \cite{us_prl} {\bf Right}: Crossover temperature
  $T^*$ as a function of $\alpha$ extracted from the data on the left
  panel by the condition $\chi_{\rm active}(T^*) = \chi_{\rm
  frozen}(T^*)$ (see text).}\label{fig_suscept} 
\end{figure}

With this picture in mind we estimate the zero frequency
susceptibility $\chi(T)$, as the sum of two contributions $\chi(T) =
\chi_{\rm active}(T) + \chi_{\rm frozen}(T)$, one arising from the
active, {\it i.e.}  non ``frozen'' spins, $\chi_{\rm active}(T)$ and
one from the ``frozen'' ones, $\chi_{\rm frozen}(T)$. In doing this,
we assume that the interaction between the frozen and the non frozen
clusters is negligible. $\chi_{\rm active}(T)$ is given by (see also
Eq. (\ref{chi2})):
\begin{eqnarray}
\chi_{\rm active}(T) = \int_0^\infty \frac{d\epsilon}{\epsilon}
\rho(\epsilon) (1-\exp{(-\beta \epsilon})) \label{suscept_nonfrozen}
\end{eqnarray} 
with $\beta = 1/T$. To estimate the density of states $\rho(\epsilon)$ using
our RG scheme we compute the distribution of the amplitudes of fields
and bonds which are decimated during the renormalization
procedure \cite{fisher-irfp}. Having computed $\rho(\epsilon)$, 
we then perform numerically the integration in
Eq. (\ref{suscept_nonfrozen}) to compute $\chi_{\rm active}(T)$. In
the Griffith's region where the restricted distribution scales with
$L$ as in Eq. (\ref{griffith_scaling}), one has $\rho(\epsilon) \propto
\epsilon^{-1+1/z}$ 
and thus $\chi_{\rm active}(T) \propto T^{-1+1/z}$. On
the other hand, each (quantum mechanically) frozen spin contributes the
susceptibility by an amount of $1/T$ and thus 
\ba
\chi_{\rm frozen}(T) = \frac{{\cal N}_{\rm frozen}(T)}{T}
\label{suscept_frozen} 
\ea
where ${\cal N}_{\rm frozen}(T)$ denotes the number of frozen spins at
temperature $T$ and its finite $T$ dependence is computed as explained
above.  We have computed $\chi(T)$ using our RG scheme for a system of
size $L=4096$ for different values of $\alpha = 0.15, 0.17, 0.19,
0.20$ and $0.22$ for $J_0=0.025$. In each case, $\chi(T)$ is averaged
over $10^4$ different realizations of the random couplings and the
plots are shown on the left panel of Fig. \ref{fig_suscept}.
Let us first consider the curves for $\alpha < 0.2$, where the
restricted distribution $\tilde P_L(\log(\Gamma_0/h))$ shows a scaling
with $L$ as in Eq. (\ref{griffith_scaling}) \cite{us_prl}. For low
temperatures still above some temperature $T^*$, $T>T^*$, one sees in
the left panel of Fig. \ref{fig_suscept} that $\chi(T)$ is dominated
by $\chi_{\rm active}(T) \propto T^{-1+1/z}$, for $\alpha < 0.2$. In
this regime of dissipation, one observes that the slope of $\chi(T)$
in a log-log plot depends on $\alpha$: this is the characteristics of
Griffith's behavior. At lower temperature $T < T^*$, $\chi(T)$ is
dominated by the $1/T$ behavior of $\chi_{\rm frozen}(T)$ coming from
the frozen clusters. Thus Fig. \ref{fig_suscept} shows that Griffith's
behavior can indeed be observed above $T^*$. For $\alpha >0.2$, the
susceptibility behaves like $\chi(T) \sim 1/T$ in the whole range of
temperature.

In addition, given that we compute separately $\chi_{\rm active}(T)$
in Eq. (\ref{suscept_nonfrozen}) and $\chi_{\rm frozen}(T)$ in
Eq. (\ref{suscept_frozen}) our numerical RG procedure allows to
estimate $T^*$ for which these two contributions are equal, 
$\chi_{\rm active}(T^*) = \chi_{\rm frozen}(T^*)$. In the right panel of
Fig. \ref{fig_suscept}, we show a plot of $T^*$ as a function of
$\alpha$. One observes in particular that $T^*$ shows an inflexion
point as the pseudo-critical point is crossed such that $T^*$ is
actually quite small in the Griffith's region.

\begin{figure}[h]
\begin{center}
\includegraphics[angle=-90,width=0.5\linewidth]{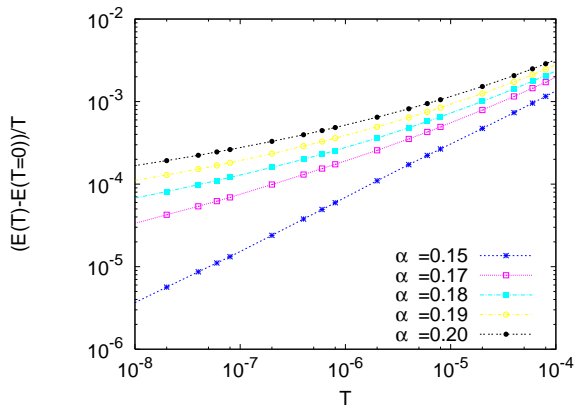}    
\caption{
  The scaled energy $T^{-1}({\cal E}(T) - {\cal E}(T=0))$, which is
  proportional to the specific heat $C_v(T)$, for the disordered chain
  coupled to an ohmic bath as a function of temperatur $T$ in a
  log-log plot. Copmputations are done using
  Eq. (\ref{int_energy}). The different curves correspond to different
  values of $\alpha$.  Note that the low temperature behavior is
  characterized by different slopes for different $\alpha$,
  corresponding to a varying exponent $1/z$. The system size is
  $L=4096$, the disorder strength $h_0=1$, $J_0 = 0.025$. For this
  value of $J_0$ the pseudo critical point is located at $\alpha=0.2$
  \cite{us_prl}.  }
\label{fig_chalspe} 
\end{center}
\end{figure}

We now turn to the specific heat $C_v(T)$ of the spin degrees of
freedom. Assuming that one can also neglect the interaction between
frozen and non frozen clusters one immediately obtains that $C_v(T) =
C_{v, {\rm active}}(T)$ given that $C_{v, {\rm frozen}}(T) =
0$. $C_v(T)$ is thus
\ba
&&C_v = \frac{\partial {\cal E}(T)}{\partial T} \label{chal_spe} \\
&&{\cal E}(T) - {\cal E}(T=0) = \frac{1}{L} \int_0^\infty \; d\epsilon
\; \rho(\epsilon)\, \epsilon \frac{\exp{(-\beta
  \epsilon)}}{1+\exp{(-\beta \epsilon)}} \label{int_energy}
\ea   
where ${\cal E}(T)$ is the internal energy at temperature $T$. In the
Griffith's region where the restricted gap distribution has a finite
$L$ scaling as in Eq. (\ref{griffith_scaling}), one expects ${\cal E}(T)
- {\cal E}(T=0) \propto T^{1+1/z}$, thus $C_v(T) \propto T^{1/z}$
without any cut-off at some temperature $T^*$. In analogy to 
$\chi(T)$ we have computed numerically ${\cal E}(T) - {\cal E}(T=0)$
(also averaged over $10^4$ disordered samples) for different values of
$\alpha$. On Fig. \ref{fig_chalspe} we show a plot of $T^{-1} ({\cal
E}(T) - {\cal E}(T=0)) \propto C_v(T)$ as a function of $T$ for
different values of $\alpha$. One observes clearly that the slope
decreases as $\alpha$ is increased, {\it i.e.} as the critical point
is reached. We tried to extract an estimate of the dynamical exponent
$z$ by fitting the curves in Fig. \ref{fig_chalspe} by $T^{-1}({\cal
  E}(T) - {\cal E}(T=0)) \propto T^{1/z}$ at low $T$ 
as well as by fitting the curves in the left panel of
Fig. \ref{fig_suscept} by $\chi(T) \propto T^{-1+1/z}$ for $T >
T^*$. Both estimates for $z$ coincide approximately  
but since the data shown are close to the pseudo-critical point (which
corresponds here to $\alpha = 0.2$), it is rather hard to
extract properly the dynamical exponent given that $1/z$  
becomes quite small, thus one would certainly need smaller
temperatures to obtain a reliable estimate of $z$.

We conclude this paragraph by noting that the data in
Fig. \ref{fig_suscept} and \ref{fig_chalspe} indicate that Griffith's behavior 
of thermodynamical quantities is observable also in the presence of
dissipation.

\subsection{Disordered ladder.}

Our previous study on Ref. \cite{us_prl} was restricted to the one
dimensional case. Here, we implement numerically the real space
renormalization defined by Eq. (\ref{decim1_j_ohm}-\ref{decim3_j_ohm})
and Eq. (\ref{decim1_h_ohm}-\ref{decim2_h_ohm}) for a disordered
ladder coupled to a dissipative bath. When considering a ladder (as
well as a two-dimensional square lattice) these decimation rules have
to be slightly modifed to take into account the topology of the system
\cite{yu-cheng_statphys}. First, Eq. (\ref{decim3_j_ohm}) has to be
modified. In this case, the two spins $2$ and $3$ are combined to a
cluster but when we compute the interactions between this cluster and
the rest of the chain, one has to consider the case in which the two
original spins $2$ and $3$ were actually coupled to the same spin
$i$. Although this does not happen in the initial ladder, such a
situation may occur during later stages of the renormalization. In
this case we set the ferromagnetic coupling of this spin $i$ with the
newly formed cluster to
\ba
\tilde J_{i,{\rm cluster}} = {\rm max}(J_{i2},J_{i3})
\;. \label{decim3_j_ladder} 
\ea   
The sum of the two bond strengths could also be taken, but does not
make a significant difference when the probability distribution of the
bond strengths is broad. 

 The decimation rule on Eq. (\ref{decim2_h_ohm}) has also to be
 modified. This rule says that 
when the spin on site $2$ is decimated, effective interactions are
generated between the neighboring sites of $2$. But during
 renormalization of the 
ladder there might already be bonds $J_{ij}$ present
between neighboring sites $i$ and $j$ of site $2$. In this case we 
replace Eq. (\ref{decim2_h_ohm}) by
\ba
\tilde J_{ij} \simeq {\rm max}(J_{ij},\frac{J_{i2} J_{2j}}{h_2}) \; . 
\ea 
The topology of the system changes drastically under
renormalization. One starts with a ladder and the decimations change
its structure into a random graph, but this change is straightforward
to implement numerically.

\begin{figure}[h]
\begin{minipage}{0.5 \linewidth}
\begin{center}
\includegraphics[angle=-90,width=\linewidth]{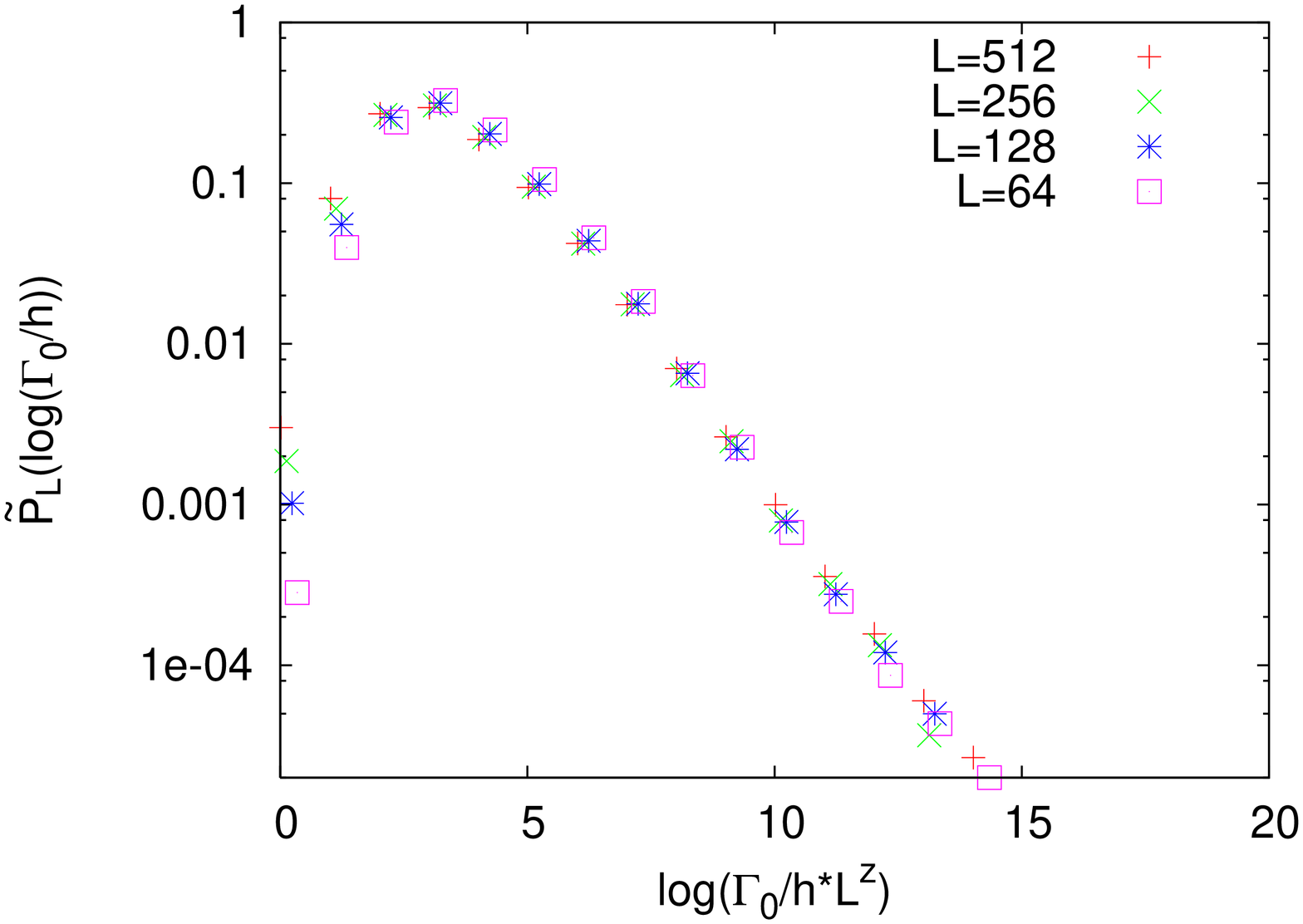} 
\end{center}
\end{minipage}\hfill
\begin{minipage}{0.5 \linewidth}
\begin{center}
\includegraphics[angle=-90,width=\linewidth]{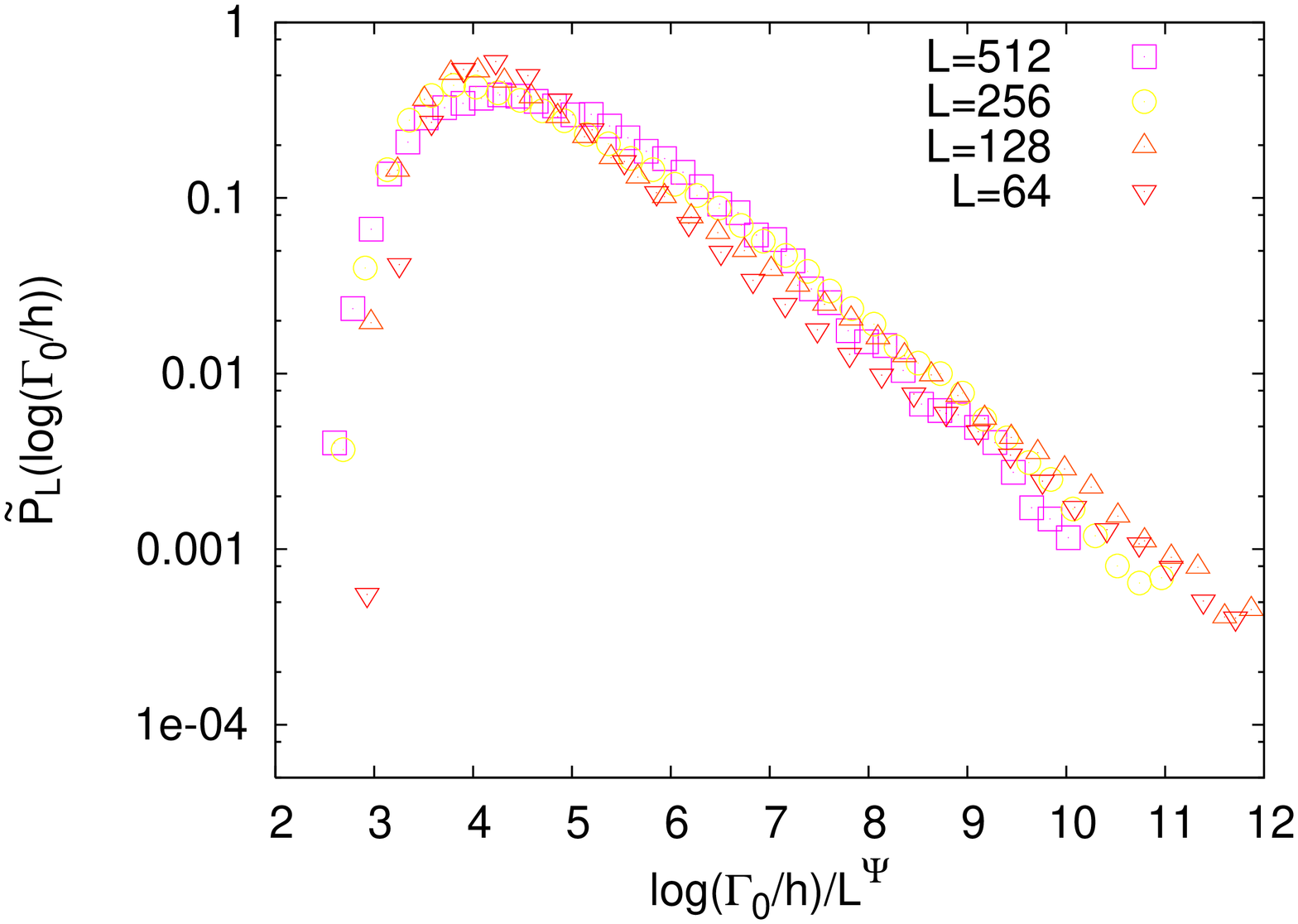} 
\end{center}
\end{minipage}
\caption{Scaling plot of the restricted distribution 
  $\tilde{P}_L(h/\Gamma_0)$ of the last field to be decimated in the
  non-frozen samples for the disordered ladder coupled to an ohmic
  bath. Here $h_0 = 1.9$, $J_0 = 0.001$.  {\bf Left}: In the
  Griffiths-region ($\alpha = 0.27$): $\tilde P_L(h/\Gamma_0)$ as a
  function of $\log(\Gamma_0/h L^z)$ for different system sizes
  $L$. The best data collapse is obtained with the dynamical exponent
  $z=1.7$. {\bf Right}: At the pseudo-critical point ($\alpha =
  0.27$): $L^\psi \tilde P_L(\Gamma_0/h)$ as a function of
  $\log{(\Gamma_0/h)}/L^\psi$ for different system sizes $L$.
  The best data collapse is obtained with the exponent 
  $\psi = 0.27(3)$.}\label{ladder_scaling1}
\end{figure}

\begin{figure}[h]
\begin{minipage}{0.5 \linewidth}
\begin{center}
\includegraphics[angle=-90,width=\linewidth]{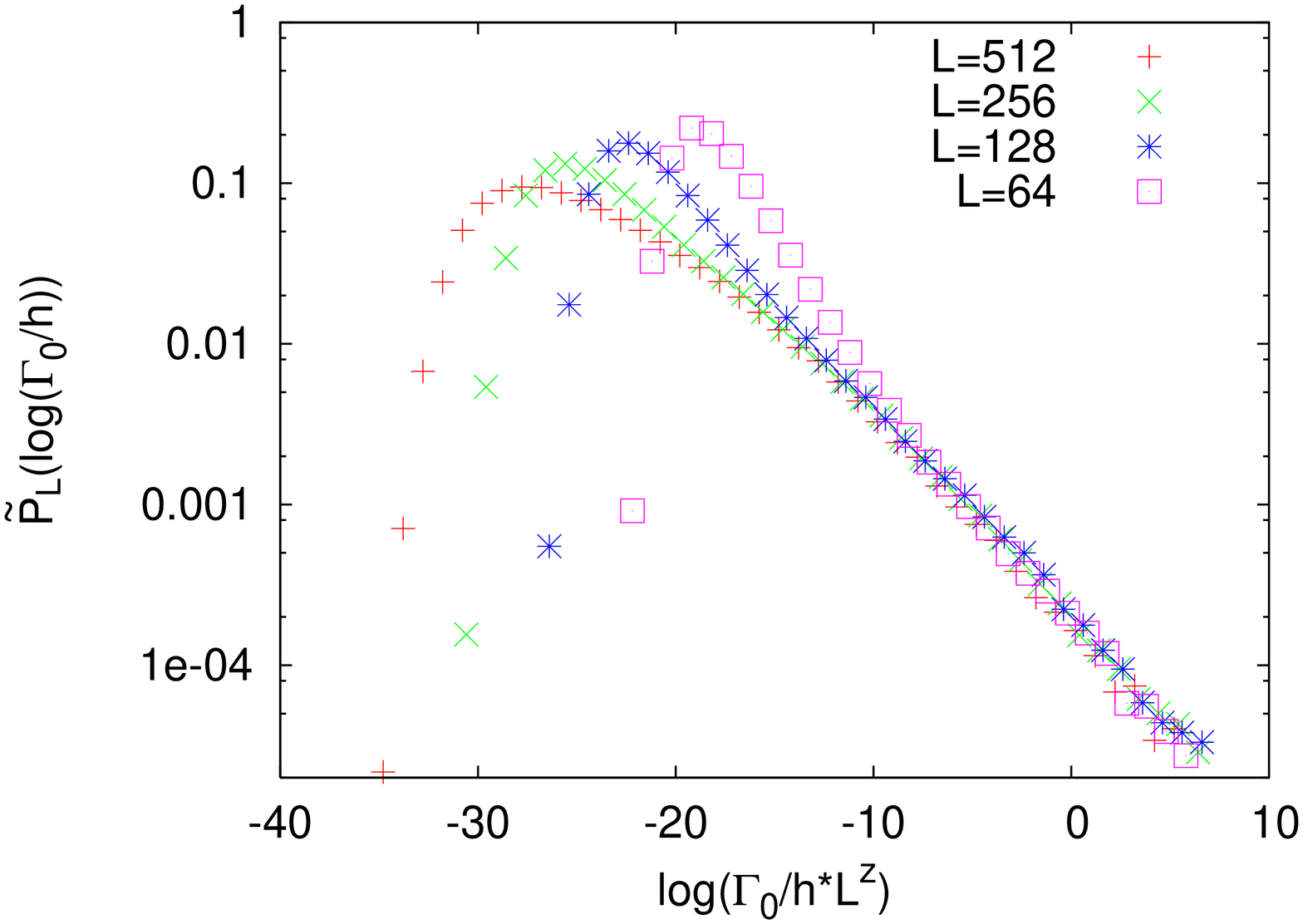} 
\end{center}
\end{minipage}\hfill
\begin{minipage}{0.5\linewidth}
\begin{center}
\includegraphics[angle=-90,width = \linewidth]{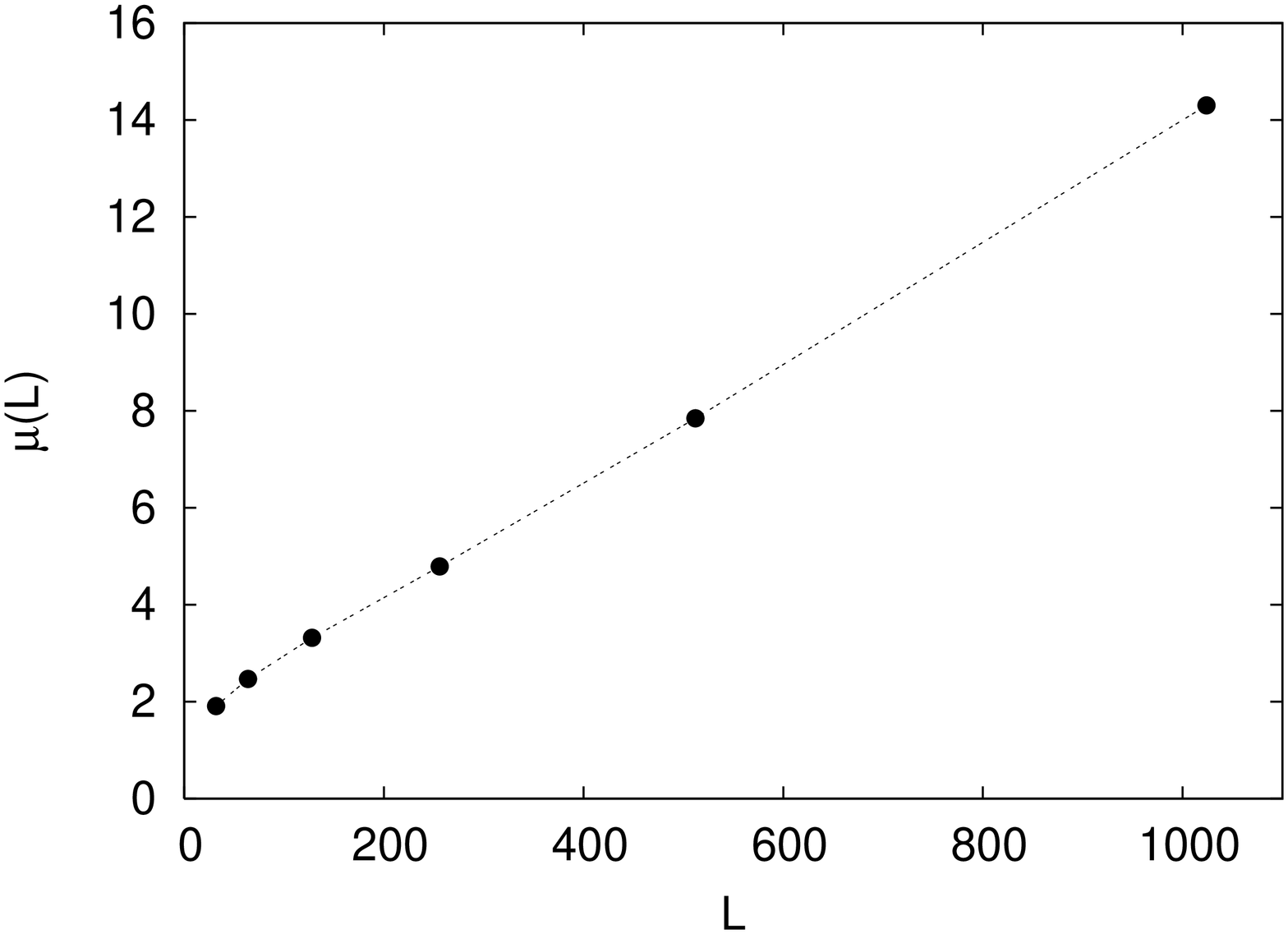} 
\end{center}
\end{minipage}
\caption{{\bf Left:} Scaling plot of the restricted distribution 
  $\tilde{P}_L(h/\Gamma_0)$ of the last field to be decimated in the
  non-frozen samples for the disordered ladder coupled to an ohmic
  bath. Here $h_0 = 1.9$, $J_0 = 0.001$ and $\alpha = 0.262$, {\it i.e.}\
  close to the pseudo-critical point at $\alpha=0.27$. The best data
  collapse is obtained by the dynamical exponent $z=7.5$.  {\bf
  Right:} Magnetic moment $\mu(L)$ as a function of $L$ in the
  Griffith's region (parameters as in the left panel). The linear
  behavior implies a non-vanishing magnetization $m_{\rm eq}$ 
  per spin.}
  \label{ladder_scaling2}
\end{figure}

In the absence of dissipation a critical point was found for $h_0 =
1.9, J_0=1$ \cite{yu-cheng_statphys}. In the following we fix $h_0 =
1.9$ and $J_0 = 0.001$ and we vary $\alpha$. As it was done previously
for the disordered chain in Ref. \cite{us_prl} we first focus on the
restricted distribution of the last fields in the samples that are non
frozen (\ref{gen_form}). For small $\alpha$, $\tilde P_L(h/\Gamma_0)$
displays Griffith's like behavior as in
Eq. (\ref{griffith_scaling}). In the left panel of
Fig. \ref{ladder_scaling1}, one plots $\tilde P_L(h/\Gamma_0)$ as a
function of $\log(\Gamma_0/h L^z)$ with $z=1.7$ for different system
sizes $L = 64, 128, 256, 512$ for $\alpha=0.2$. The good data collapse
of the curves for different $L$ is in a good agreement with Griffith's
scaling (\ref{griffith_scaling}).
We observe that the dynamical exponent $z$ increases with increasing
$\alpha$. This is depicted in the left panel of
Fig. \ref{ladder_scaling2} where one plots again $\tilde
P_L(h/\Gamma_0)$ as a function of $\log(\Gamma_0/h L^z)$ for different
system sizes $L$ but with $z=7.5$ for $\alpha = 0.262$.
However, despite the fact that the gap distribution displays
Griffith's behavior, the magnetization is already finite. This can be
seen by computing the magnetic moment of the last remaining cluster as
a function of the system size $L$, see the right panel of
Fig. \ref{ladder_scaling2}. This behavior, which is due to frozen
clusters is very similar to the one observed for the disordered chain
\cite{us_prl}.   

Finally, one reaches a value of $\alpha$ where $z$ is diverging
and one observes a scaling characteristic for an infinite randomness
fixed point as in Eq. (\ref{irfp_scaling}). This is shown on the right
panel of  
Fig. \ref{ladder_scaling1} where we plot $L^\psi \tilde
P_L(\Gamma_0/h)$ as a function of $\log{(\Gamma_0/h)}/L^\psi$ with
$\psi = 0.27(3)$.

\subsection{Two-dimensional square lattice.}

\begin{figure}
\begin{minipage}{0.5\linewidth}
\begin{center}
\includegraphics[angle=-90,width=\linewidth]{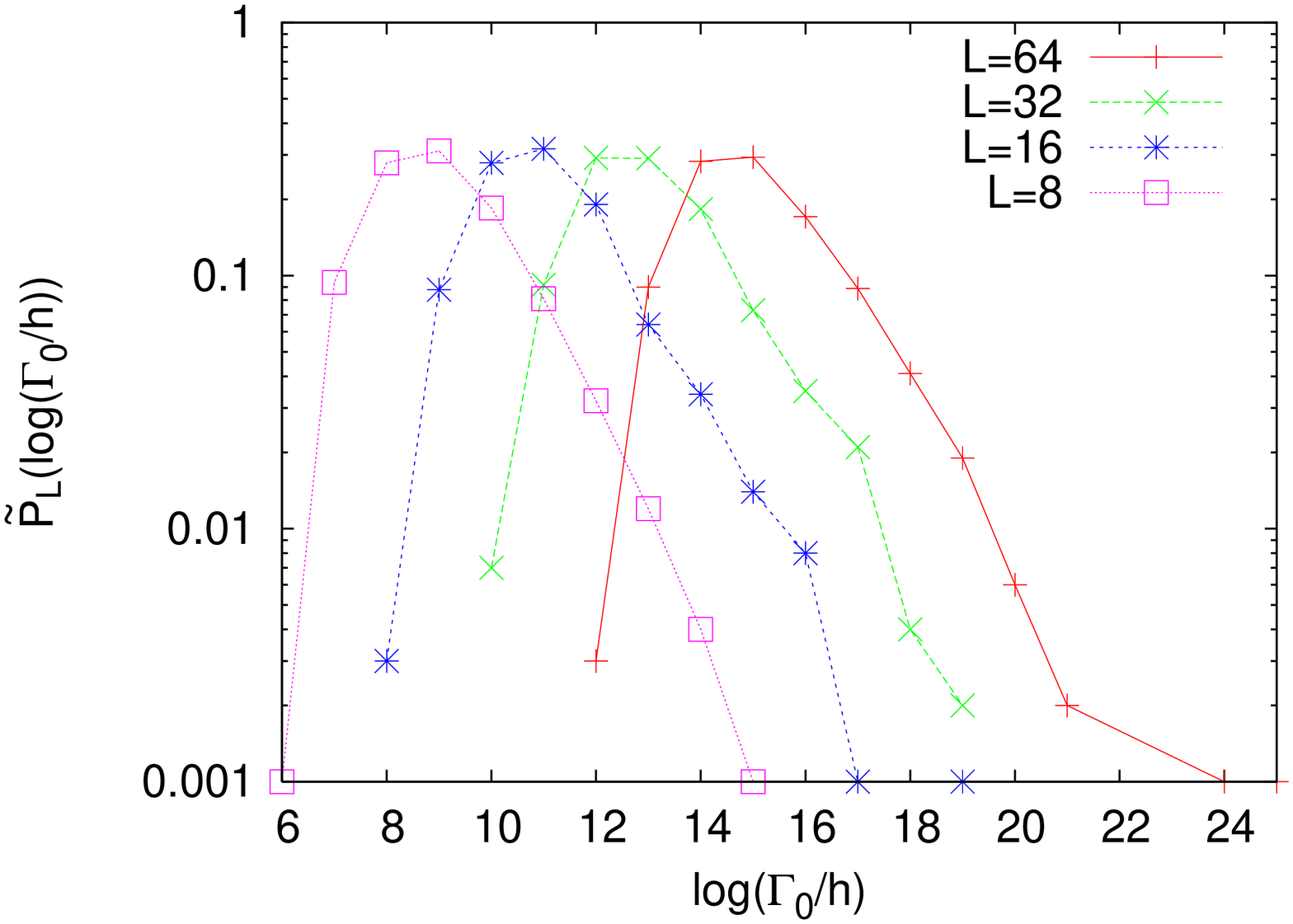}  
\end{center}
\end{minipage}\hfill
\begin{minipage}{0.5\linewidth}
\begin{center}
\includegraphics[angle=-90,width=\linewidth]{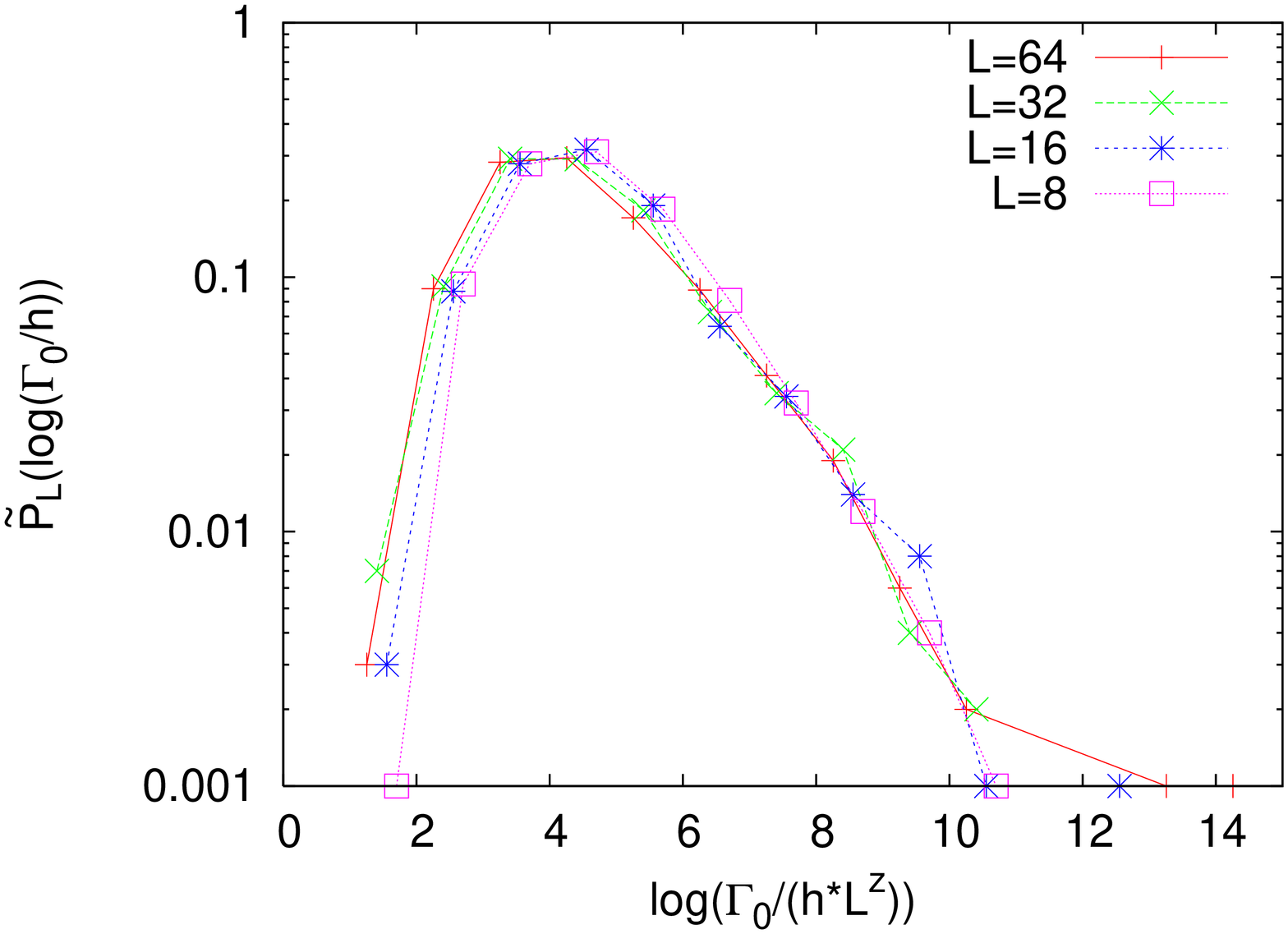}  
\end{center}
\end{minipage}
\caption{The restricted distribution 
  $\tilde{P}_L(h/\Gamma_0)$ of the last field to be decimated in the
  non-frozen samples for the disordered square lattice coupled to an
  ohmic bath. Parameter values fall into the Griffiths-region: the
  disorder strength is $h_0 = 5.35$, $J_0 = 0.0001$ and the coupling
  to the bath is $\alpha=0.3$. {\bf Left}: $\tilde
  P_L(\log(\Gamma_0/h))$ as a function of $\log{(\Gamma_0/h)}$ for
  different system sizes $L$. {\bf Right}: Scaling plot of the data in
  the left panel: $\tilde P_L(\log(\Gamma_0/h))$ as a function of
  $\log(\Gamma_0/h L^z)$ for different system size $L$. The best data
  collapse is obtained with $z=3.1$.}\label{2d_scaling_griff} 
\end{figure}

\begin{figure}[h]
\begin{minipage}{0.5\linewidth}
\begin{center}
\includegraphics[angle=-90,width=\linewidth]{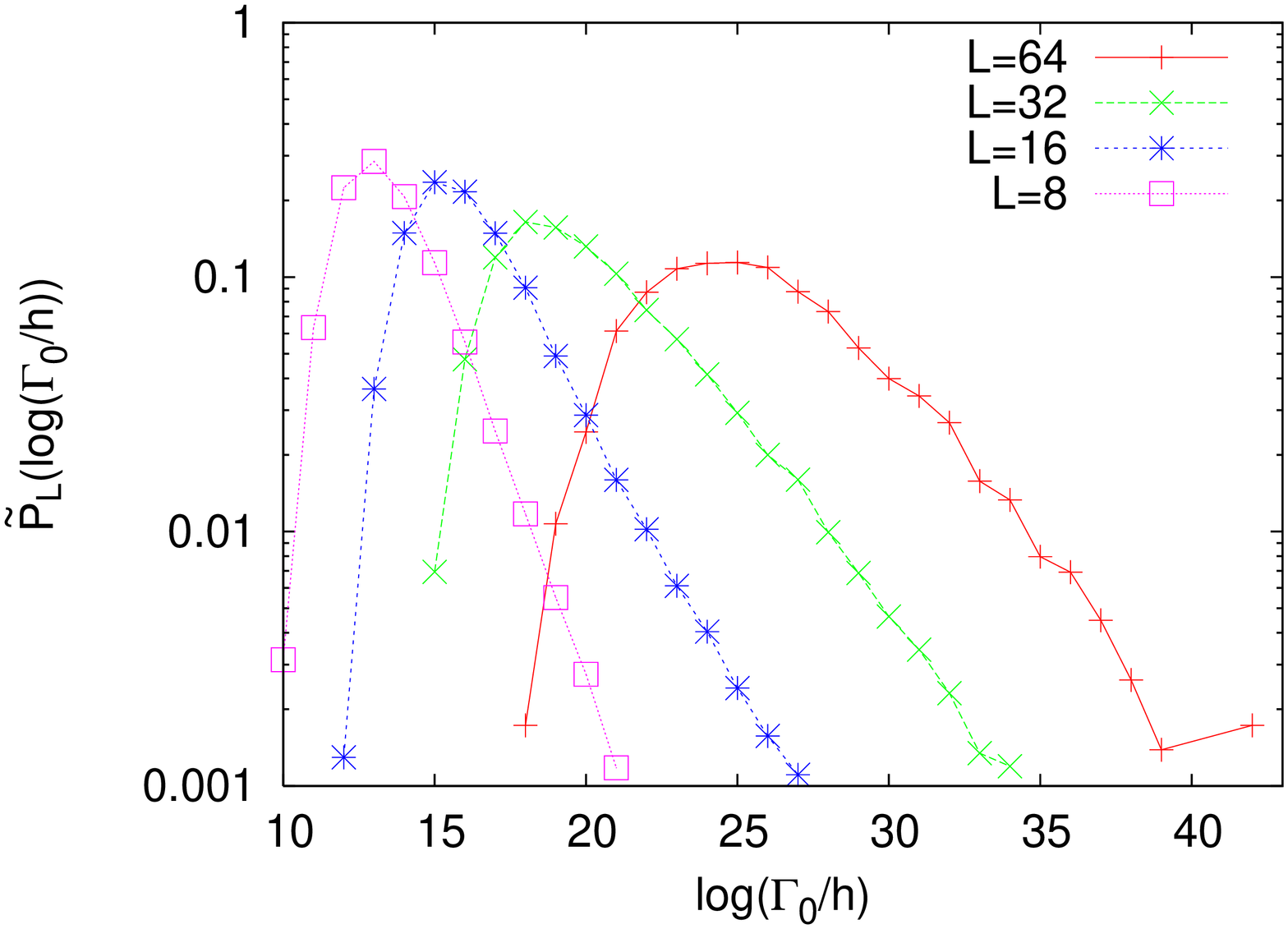}  
\end{center}
\end{minipage}\hfill
\begin{minipage}{0.5\linewidth}
\begin{center}
\includegraphics[angle=-90,width=\linewidth]{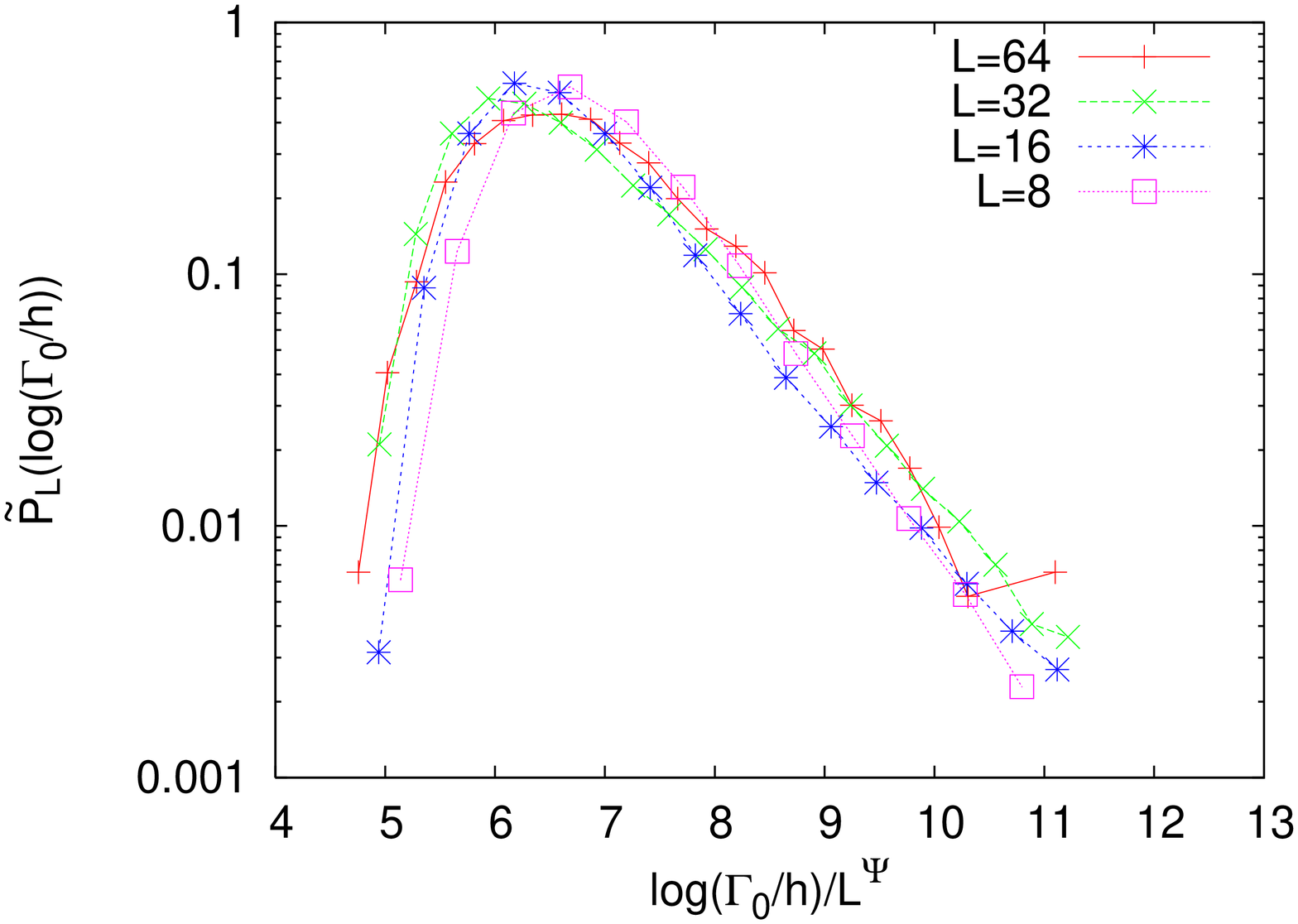}  
\end{center}
\end{minipage}
\caption{The restricted distribution 
  $\tilde{P}_L(h/\Gamma_0)$ of the last field to be decimated in the
  non-frozen samples for the disordered square lattice coupled to an
  ohmic bath. Parameter values correspond to the pseudo-critical
  point: the disorder strength is $h_0 = 5.35$, $J_0 = 0.0001$ and the
  coupling to the bath is $\alpha=0.37$. {\bf Left}: $\tilde
  P_L(\log(\Gamma_0/h))$ as a function of $\log{(\Gamma_0/h)}$ for
  different system size $L$. {\bf Right}: Scaling plot of the data in
  the left panel: $\tilde L^\psi P_L(\log(\Gamma_0/h))$ as a function
  of $\log(\Gamma_0/h)/L^\psi$. The best data collapse is obtained
  with the exponent $\psi=0.32$.}\label{2d_scaling_irfp}
\end{figure}

We have also implemented the decimation rules in two dimensions for a
square lattice. Here also the topology of the system changes
drastically during renormalization.  In the absence of dissipation a
critical point was found for $h_0 = 5.35, J_0=1$. Here we include
dissipation, fix $J_0 = 0.0001$ and vary $\alpha$. At small $\alpha$
one observes Griffith's like behavior of the restricted distribution
as in Eq. (\ref{griffith_scaling}). In the left panel of
Fig. \ref{2d_scaling_griff}, we plot $\tilde P_L(\log(\Gamma_0/h))$ as
a function of $\log{(\Gamma_0/h)}$ for different system sizes $L=8,
16, 32, 64$ and $\alpha =0.3$. On the right panel, we show that these
curves for different fall $L$ on a master curve if one plots them as a
function of $\log(\Gamma_0/h L^z)$ with $z=3.1$.
As we increase the value of $\alpha$, one observes that $z$ is also
increasing and eventually we identify a pseudo critical point, here
for $\alpha=0.37$, where the 
restricted distribution has a scaling form characteristic for an IRFP
as in Eq. (\ref{irfp_scaling}) with $\psi = 0.32$. This is shown in
Fig. \ref{2d_scaling_irfp}.   

\section{Super-ohmic dissipation.}

We have implemented numerically the decimation rules for the
super-ohmic bath, which corresponds to $s>1$. In this case the
amplitude ${\cal A}$ in Eq. (\ref{eq_amplitudeA}) and ${\cal A}'$ in
Eq. (\ref{eq_amplitudeAprime}) which enter the decimation rules are
given by
\ba
{\cal A} = \exp{\left(-\frac{\alpha_2}{s-1}\left( 1 -
  \left(\frac{pJ_2}{\Omega_2}\right)^{s-1}  \right)
  -\frac{\alpha_3}{s-1}\left( 1 - 
  \left(\frac{pJ_2}{\Omega_3}\right)^{s-1}  \right)     \right)} \; ,
\nonumber \\
{\cal A}' = \exp{\left(-\frac{\alpha_2}{s-1}\left(1 -
  \left(\frac{ph_2}{\Omega_2}\right)^{s-1}   \right) \right)}  \;.
\ea
For $s>1$ iterations of the decimation rules (\ref{decim1_h_ohm})
always converge to 
a fixed point value $h_2^* > 0$ given in
Eq. (\ref{h_2star_explicit}). Consequently, the spins can not be
frozen by the dissipative bath.

We first present results for $s=3$, which corresponds to a phonon
bath, and one fixes the coupling to te bath to $\alpha = 0.5$ and the
strength of the random transverse field to $h_0=1.0$. All data
presented here were obtained by averaging over $10^4$ different
realizations of the disordered couplings.

\begin{figure}[h]
\begin{center}
\includegraphics[angle=-90,width=0.5\linewidth]{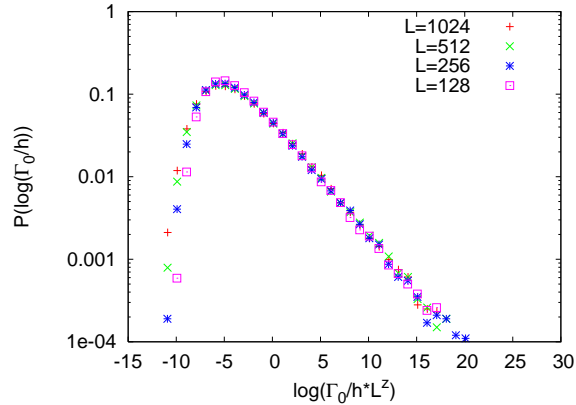}
\caption{Scaling plot of the probability distribution of the last 
decimated field for the disordered chain with super-ohmic dissipation,
here $s=3$: $P_L(\Gamma_0/h)$ vs. $\log{(\Gamma_0/h L^z)}$ (scaling in
the Griffiths region) for different system size $L$. The best data
collapse is obtained with $z=2.87$. The disorder strength is $h_0 =
1.0$ and $J_0 = 0.55$ and the coupling to the bath is $\alpha =
0.5$.}\label{super_ohm_griff}
\end{center}
\end{figure}

\begin{figure}[h]
\begin{minipage}{0.5 \linewidth}
\begin{center}
\includegraphics[angle=-90,width=\linewidth]{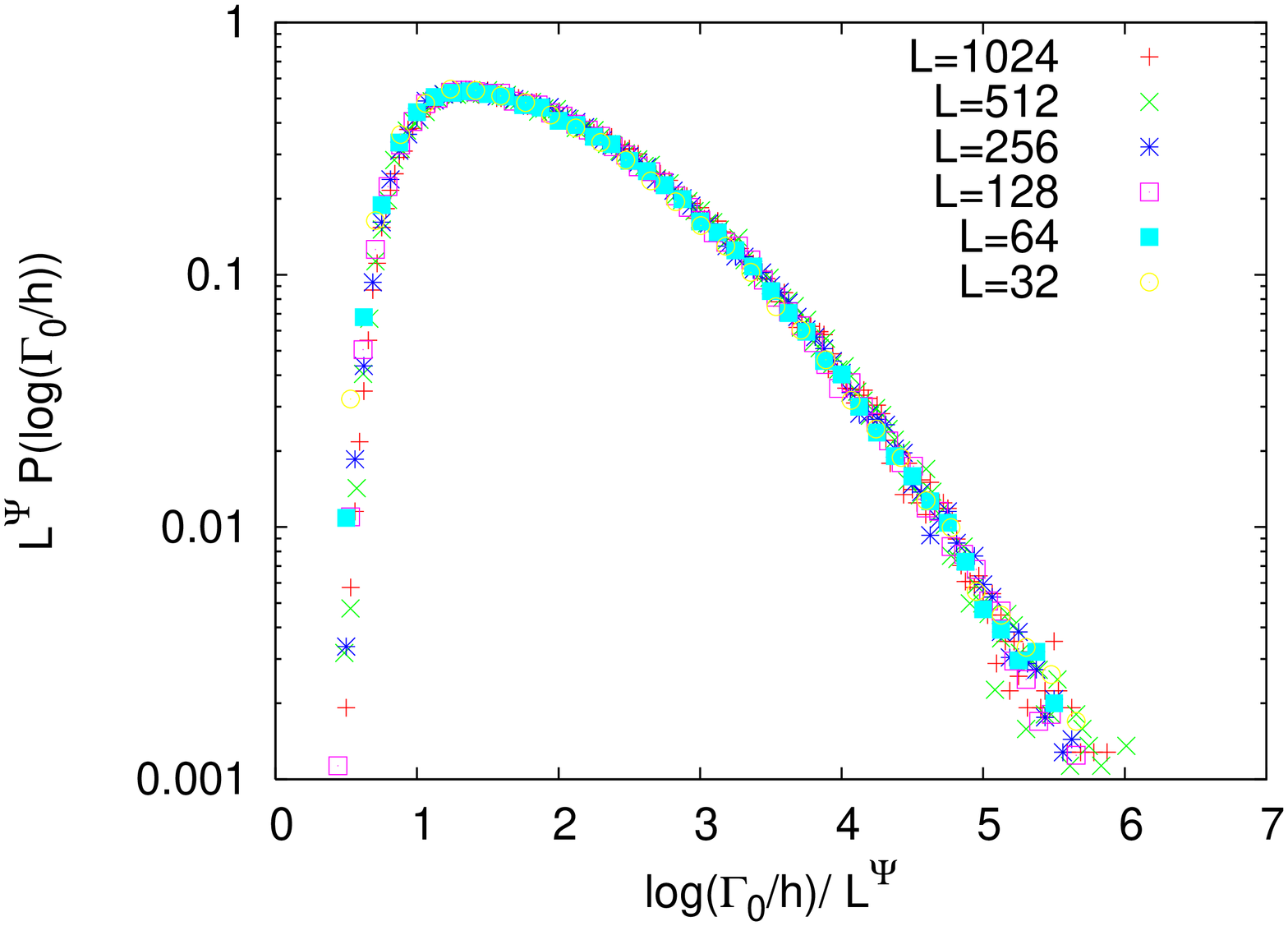} 
\end{center}
\end{minipage}\hfill
\begin{minipage}{0.5 \linewidth}
\begin{center}
\includegraphics[angle=-90,width=\linewidth]{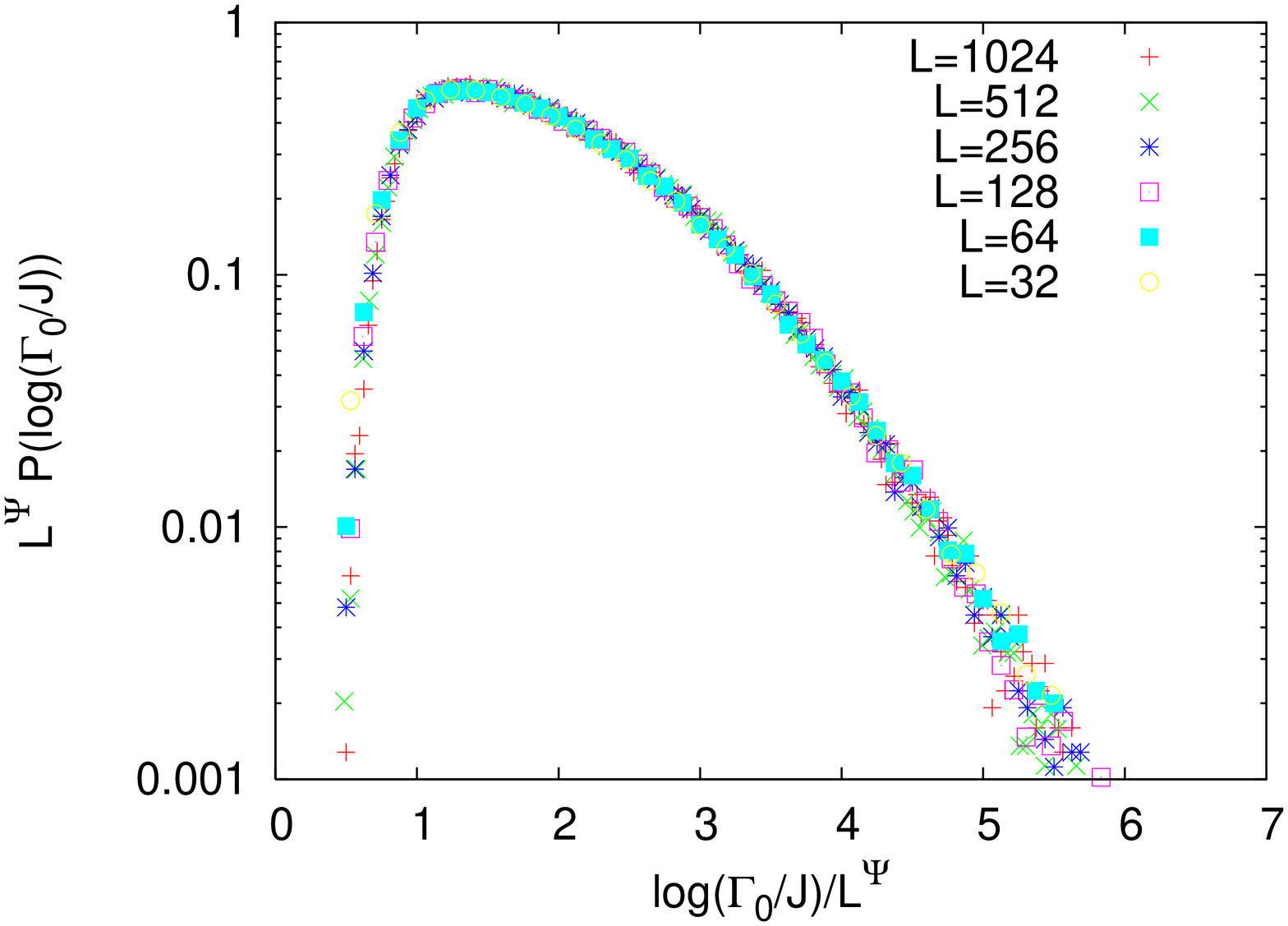} 
\end{center}
\end{minipage}
\caption{{\bf Left:} Scaling plot of the probability distribution of the last 
  decimated field for the disordered chain with super-ohmic
  dissipation, here $s=3$: $L^\psi P_L(\log(\Gamma_0/h))$ vs.
  $(\log(\Gamma_0/h))/L^\psi$ (IFRP scaling) for different system size
  $L$ with $\psi = 1/2$. The disorder strength is $h_0 = 1.0$ and $J_0
  = 0.55$ and the coupling to the bath is $\alpha = 0.78$.  {\bf
  Right:} The same as in the left panel but for the bond distribution
  instead of the field distribution. Note the similarity of the
  two distributions.}\label{super_ohm_irfp} 
\end{figure}

We first focus on low value of $J_0$. In Fig. \ref{super_ohm_griff},
one shows a plot of $P_L(\Gamma_0/h)$, the distribution of the
transverse field acting on the last remaining cluster as a function of
$\log(\Gamma_0/h L^z)$ for different system sizes with $z=2.87$. The
good data collapse of these different curves suggests that
$P_L(h/\Gamma_0)$ exhibits Griffith's behavior:
\begin{eqnarray}
P_L(\log(\Gamma_0/h)) = {\cal P}(\log(\Gamma_0/h L^z))\;.
\label{griffith_scaling_superohm}
\end{eqnarray}
Notice that, at variance with the case of ohmic dissipation
(\ref{gen_form}) one has here $A_L=1$.
If one increases $J_0$, $z$ is also increasing and for some critical value of $J_0=J_{0c}$,
here $J_{0c}=0.78$ one observes a scaling characteristic for an IRFP
\begin{eqnarray}
P_L(\log(\Gamma_0/h)) = 
L^{-\psi}{\cal P}_{\rm{IRFP}}(L^{-\psi}\log(\Gamma_0/h))\;,
\label{super_ohm_irfp_scaling} 
\end{eqnarray}
with $\psi=1/2$ as in the case without dissipation \cite{fisher}. This
is shown in the left panel of  
Fig. \ref{super_ohm_irfp}. 
In the absence of dissipation random fields and random bonds play a
symmetric role in the RTFIC. This is in principle not the case when
one includes dissipation in the Hamiltonian (\ref{Def_H}). However,
this symmetry is restored asymptotically, close to the critical
point. To show this, we have computed $P_L(J/\Gamma_0)$ where $J$ is
the last decimated bond. In Fig. \ref{super_ohm_irfp} we show a plot
of $L^\psi P_L(\log(\Gamma_0/J))$ as a function of
$(L^{-\psi}\log(\Gamma_0/h))$ with $\psi = 1/2$ for $J_0=J_{0c}$. The
good data collapse, together with the similarities between the plots
shown on both panels of Fig. \ref{super_ohm_irfp} suggest indeed that
this symmetry between bonds and fields is restored at the critical
point.

\begin{figure}[h]
\begin{center}
\includegraphics[angle=-90,width=0.5\linewidth]
{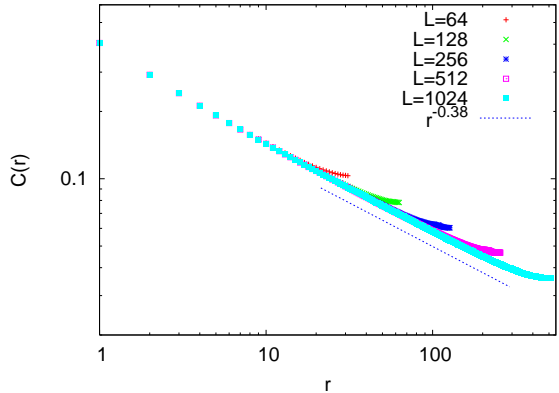}
\caption{The disorder averaged correlation function $C(r)$ at the
  critical point for the chain coupled to a super-ohmic bath (here
  $s=3$) for different system size $L$. The disorder strength
  is $h_0 = 1$ and $J_0 =0.78$ and the coupling to the bath is $\alpha
  = 0.5$. The error bars are smaller than the size of the symbols. The
  decay exponent $\simeq 0.38$ gives $\phi \psi \simeq 0.81$ as in the
  case without dissipation \cite{fisher}.}\label{Fig_correl}
\end{center}
\end{figure}

To characterize this IRFP, we have also computed the combinations of
the products of the exponents $\varphi \psi$ where $\varphi$ is
another independent exponent associated to this IRFP. This can be
measured by computing the disorder averaged correlation function
$C(r)$ at the transition.  We compute it by keeping track of the
clusters during the decimation procedure and compute $C(r) = L^{-1}
\sum_{i} w_{i,i+r}$ where $w_{i,j} = 1$ if the sites $i$ and $j$
belong to the same cluster, and $w_{i,j} = 0$ otherwise. We have
checked that for RTFIC without dissipation at the critical point this
gives the correct exponent \cite{fisher} within $5\%$ accuracy. A plot
of $C(r)$ is shown on Fig. \ref{Fig_correl} for different system sizes
$L=64, 128, 256, 512$ and $1024$. This plot shows that $C(r) \propto
r^{-\eta}$ with $\eta = 0.38(1)$, as in the case without
dissipation~\cite{fisher}. 

\begin{figure}[h]
\begin{center}
\includegraphics[angle=-90,width=0.5\linewidth]{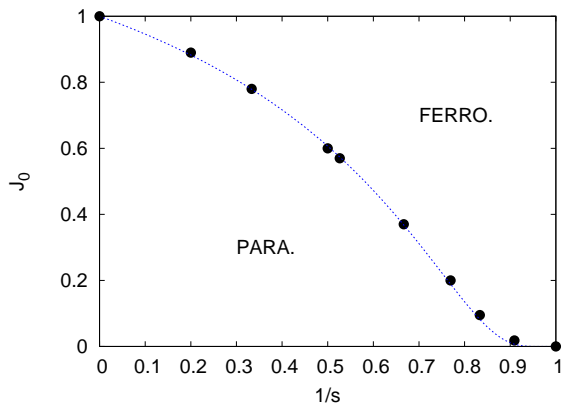}
\caption{  
Phase diagram for a disordered chain coupled to a super-ohmic
  bath (except for $s=1$) and $h_0=1, \alpha=0.5$. The error bars are smaller than the size of the symbols. Along the line the
  critical behavior is governed by an IRFP with $\psi=1/2$ (see
  Fig. \ref{super_ohm_irfp} above). 
 The dashed line is the
exact expression of the critical coupling given in the text
 (\ref{critical_super}).}\label{ph_diag} 
\end{center}
\end{figure}

We have repeated the same procedure for different values of $s$ and found
the critical value $J_{0c}(s)$.  We thus obtain the phase
diagram in the plane $(J_0,1/s)$ shown in Fig. \ref{ph_diag} where a
critical line separates a paramagnetic phase from a ferromagnetic one.
Along this line, we have found a scaling like in
Eq. (\ref{super_ohm_irfp_scaling}) with an exponent $\psi=1/2$,
independently of $s$. One can actually estimate the shape of the
critical line in Fig. \ref{ph_diag} by assuming that the main effect
of dissipation is to reduce the amplitude of the random transverse
field $h_0$ to $h_0^*$ given by Eq. (\ref{h_2star_explicit}). If one
further assumes that bonds and fields play a symmetric role at the
critical point (which is fully compatible with our numerical results
in Fig. \ref{super_ohm_irfp}), the critical point is then given by the
relation $J_0 = h_0^*$, as in the RTFIC \cite{fisher}. This leads to
\begin{eqnarray}
J_{0c} = h_0 \exp{\left(\frac{-\alpha}{s-1}\right)} \;,
\label{critical_super} 
\end{eqnarray}
which is actually in very good agreement with our numerical estimates
for the critical line in Fig. \ref{ph_diag}. Using the same arguments,
one can also derive analytically the behavior of the dynamical
exponent when approaching the critical point, this yields
\begin{eqnarray}
z \propto \frac{1}{2\delta} \quad, \quad \delta = \frac{1}{2}
\log{\left( \frac{h_0 e^{-\alpha/(s-1)}}{J_0}\right)} \;,
\end{eqnarray}
which we have checked to be in good agreement with numerical results. 

Our results thus suggest that for $s>1$, the large scale $L > L_s$
properties of the system with super ohmic dissipation behaves at
criticality as the dissipationless system. On the other hand, one
expects that $L_s$ is diverging when $s \to 1$. To estimate its
behavior close to $s=1$ one observes that the typical energy scale at
criticality is given by $E_s \propto h_0 \exp(-\alpha/(s-1))$
(\ref{critical_super}).  But given that the critical behavior is
governed by an IRFP, one expects that $L_s \propto (\log
E_s)^{1/\psi}$, with $\psi = 1/2$, see
Fig. \ref{super_ohm_irfp}. Therefore we estimate
\ba
L_s \propto \frac{\alpha^2}{(s-1)^2} \;,
\ea
the length above which the system with super ohmic dissipation behaves
like the one without dissipation.

\section{Conclusion}

In this paper we have developed a real space renormalization, which
combines the SDRG for strongly disordered quantum magnets with the
adiabatic renormalization for spin-boson systems, to study disordered,
ferromagnetically interacting transverse Ising systems coupled to a
dissipative bath. In the important case of ohmic dissipation, we have
first extended our previous study in Ref. \cite{us_prl} to describe
thermodynamical properties. In particular we have shown that
Griffith's Mc-Coy singularities are visible in the spin specific heat
at all temperatures and in the magnetic susceptibility above a (small)
temperature $T^*$. For weak dissipation this temperature is extremely
small and system sizes above which classical behavior in the
susceptibility becomes visible are extremely large, which represents a
major obstacle for numerical studies \cite{cugliandolo}.

We have also shown that the disordered ladder as well as the $2d$
disordered square lattice coupled to a ohmic bath displays the same
behavior.  Using this real space renormalization, we also studied
the case of super-ohmic dissipation ($s>1$). There we have found a
quantum phase transition described by an IRFP, which is the same as
the one found without dissipation. Such a scenario is expected to hold
also in higher dimensions.

It would be natural to extend this approach to sub-ohmic dissipation
($s<1$). Unfortunately, it is well known that in that case the adiabatic
renormalization fails to describe correctly the single spin-boson, which in
itself has been the subject of recent works \cite{vojta_subohmic}. Therefore
the problem of an infinite chain (possibly disordered) coupled to a sub-ohmic
bath remains a challenging problem which surely deserves further
investigations.

A final remark concerns the effect of dissipation upon magnetic
systems with a continuous symmetry instead of the discrete (Ising) case
we studied in this work. Griffiths-McCoy singularities are much weaker
in systems with a continuous symmetry \cite{sachdev}, and one would
therefore expect that coupling to a dissipative bath would {\it not}
freeze the strongly coupled regions, but enhance their singular behavior.
What actually happens can elegantly be classified according to whether
rare regions including their long range interactions in imaginary time
due to dissipation are below, at or above their upper critical dimension
\cite{vojta-rev}. A disordered itinerant antiferromagnet, for instance,
was recently studied with the strong disorder renormalization group
and an infinite randomness fixed point was found \cite{hoyos},
including the accompanying algebraic Griffiths-McCoy singularities.
On the other hand non-intinerant antiferromagnets, involving localized
magnetic moments, in spatial dimensions larger than 2 like the
Heisenberg antiferromagnet on the square lattice, will not show
pronounced Griffiths-McCoy behavior since here the Ne\'el ordered
ground state is very robust against disorder \cite{laflo} and no
quantum critical point occurs. The effect of dissipation upon strongly
disordered magnets thus depends crucially on the effect of 
disorder itself on the system's ground state.

\ack{We thank Y.C. Lin for useful discussions and acknowledge financial 
support of the Deutsche Forschungsgemeinschaft (DFG).}

\begin{appendix}

\section{A toy model for an Ising chain with ohmic
  dissipation}\label{rtic_diluted}

To understand qualitatively the full problem described by the
Hamiltonian (\ref{Def_RTFIC}) with ohmic dissipation, it is
instructive to consider a simpler model where one considers a RTFIC
without dissipation but with a finite 
fraction $\rho$ of zero transverse fields. 
We thus study in detail in this appendix the RTFIC Hamiltonian with $k$
sites having zero transverse fields ($\rho = k/L$)
\be
{\cal H}=-\sum_{i=1}^L J_i\si\sj+\sum_{i=1}^L h_i\sigma_i^x
\qquad{\rm and}\quad
h_{i_1}=\ldots=h_{i_k}=0\;. \label{h_app}
\ee
First, one immediately sees that the distribution $P_L(h/\Gamma_0)$
shows the same behavior as in Eq. (\ref{gen_form}) with $A_L \sim
e^{-L/L^*}$ and in the small $\rho$ limit, $L^* \propto
\rho^{-1}$. Besides, the local zero frequency susceptibility is  
\ba
\chi_i(\omega=0)&=&
\int_0^\beta d\tau\langle\si(\tau)\si(0)\rangle\\
&=&
\int_0^\beta d\tau \frac{1}{\cal Z} {\rm Sp}\{
\rho e^{{\cal H}\tau}\si e^{-{\cal H}\tau}\si\} \nonumber \\
&=&
\frac{1}{\cal Z}
\sum_{\{n,m\;(E_n\ne E_m)\}}
\frac{e^{-\beta E_m}-e^{-\beta E_n}}{E_n-E_m}
\vert\langle n\vert\si\vert m\rangle\vert^2 \nonumber \\
&+& \frac{1}{\cal Z}
\sum_{\{n,m\;(E_n= E_m)\}}
\beta e^{-\beta E_n}
\vert\langle n\vert\si\vert m\rangle\vert^2 \;,
\label{chi}
\ea
where $\{|n\rangle\}$ is a complete basis of eigenvectors of ${\cal
  H}$ (\ref{h_app}). Their corresponding eigenvalues $E_n$ are such
  that $E_0 < E_1 < E_2 < ...$. The first
term in Eq. (\ref{chi}) yields at zero temperature in the non-degenerate case 
(all transverse fields positive, finite system size $L$) the known formula 
\be
\chi_i^{T=0}(\omega=0)=2\sum_{n\ne0}
\frac{\vert\langle n\vert\si\vert 0\rangle\vert^2}{E_n-E_0} \; ,
\ee
since then $\langle 0\vert\si\vert0\rangle=0$ and the last term in
  Eq. (\ref{chi}) vanishes.

If one or more transverse fields vanish the Hamiltonian becomes
block-diagonal. We choose $z$-representation, such that states can be
denoted $\psi=|S_1,\ldots,S_L\rangle$, with $S_i=\pm1$. For
convenience we
permute the components such that the site $i_1,\ldots,i_k$ with the
$k$ vanishing transverse fields $h_{i_1},\ldots,h_{i_k}$ stand to the
left: $\psi=|S_{i_1},\ldots,S_{i_k},S_{j_1},\ldots,S_{j_{L-k}}\rangle$. All
$2^k$ blocks are identical up to the diagonal part $-\sum_i J_i S_i
S_{i+1}$. As a result the two blocks belonging to the states with
$S_{i_1}=\ldots=S_{i_k}$ (ferromagnetically aligned ``frozen'' spins),
have the lowest ground state energy. Obviously
\ba
\langle\psi \vert\sigma_{i_p}^z\vert
\psi\rangle&=&S_{i_p}\quad{\rm for}\quad
p=1,\ldots,k\label{note}\\
\langle\psi\vert\sigma_i^z\vert
\psi'\rangle&=&0\quad{\rm for}\quad
i=1,\ldots,L,\\
&&\quad{\rm and}\quad
(S_{i_1},\ldots,S_{i_k})\ne(S_{i_1}',\ldots,S_{i_k}') \nonumber \;.
\ea
At low temperatures ($T\to0$) the main contributions in the sums in
(\ref{chi}) comes form the terms with $E_n=E_0$ or $E_m=E_0$, the
ground state energy. When $k\ne0$ there are two ground states $\vert
0_S\rangle$, one with $S=+1$, one with $S=-1$. For $T\to0$ ${\cal Z}$
can be replaced by $2e^{-\beta E_0}$, since ${\cal Z}=2e^{-\beta
E_0}(1+\sum_m e^{-\beta(E_m-E_0)})$.  The two ground states produce
also an extra factor 2 (in addition to the one for the sum over $n\ne m$,
where either $\vert n\rangle$ or $\vert m\rangle$ can be the ground
state):
\ba
\hspace*{-2cm}\chi_i(\omega=0)&=&
\frac{4}{2e^{-\beta E_0}}
\sum_{n\ne0} 
\frac{e^{-\beta E_0}-e^{-\beta E_n}}{E_n-E_0}
\vert\langle n\vert\si\vert 0\rangle\vert^2
+ 
\frac{2}{2e^{-\beta E_0}}
\,\beta\,e^{-\beta E_0}
\vert\langle 0\vert\si\vert 0\rangle\vert^2 \nonumber \\
&=&
2\sum_{n\ne0}\frac{1-e^{-\beta(E_n-E_0)}}{E_n-E_0}
\vert\langle n\vert\si\vert 0\rangle\vert^2
+\beta\vert\langle 0\vert\si\vert 0\rangle\vert^2 \;.
\label{chi2}
\ea
The usual argument leading to $\chi_i(\omega=0)\propto T^{-1+1/z}$ in
the Griffiths-McCoy phase of the RTFIC with $k=0$ involves neglecting
the terms $n>1$ in the first sum in (\ref{chi2}).  This leads to
$\chi_i(\omega=0)\sim(\Delta E)^{-1}$, where $\Delta E=E_1-E_0$ is the
gap, which follows the distribution $P(\Delta)\sim\Delta^{-1+1/z}$. In
the present case this distribution has a cut-off at $\Delta_{\rm 
min}$ that is exponentially small in $L^*$, the average distance
between sites with zero transverse fields. Thus one expects the first
term of (\ref{chi2}) to display $T^{-1+1/z}$-behavior down to a
temperature $T_{\rm min}=\Delta_{\rm min}^{z/(z-1)}\sim
e^{-aL^*}$. The second term is $\beta\cdot m_i^2$, where $m_i$ is the
local 
magnetization in (one of) the ground states - and is non-zero due to 
(\ref{note}). It decays exponentially with the distance $x$ from the nearest 
frozen site $i_p$: $m_i~e^{-x/2}$, thus the average over all sites is 
approximately
\be
[m_i^2]_{\rm av}\sim\frac{1}{L^*/2}\int_0^{L^*/2}dx\,e^{-x}\sim\frac{1}{L^*}.
\ee
Thus a $T^{-1}$-behavior coming from the second term in (\ref{chi2})
with amplitude of order $1/L^*$ competes with a $T^{-1+1/z}$-behavior
with amplitude of order $1$ coming from the first term in
(\ref{chi2}). The latter dominates for temperatures above a cross-over
temperature $T_{\rm cross}$, which is given by
\be
T_{\rm cross}\sim (L^*)^{-z}\;,
\ee
which is larger then $T_{\rm min}$ (caused by the finite average
length $L^*$ of the segments) but still very small when $L^*\gg1$
(for instance for $L^*=10^3$ and $z=2$ one has $T_{\rm cross}\sim
10^{-6})$.


\end{appendix}

\end{document}